\documentclass[11pt,a4paper]{article}
\usepackage{jcappub}
\usepackage{amssymb,amsmath,amsfonts}
\usepackage{bm}
\usepackage{amsfonts}
\usepackage{latexsym}
\usepackage{graphicx}
\usepackage{amsmath}
\usepackage{palatino}
\usepackage{mathpazo}
\usepackage{textcomp}
\usepackage{subfigure}
\usepackage{float}
\usepackage{booktabs}
\usepackage{mathtools}
\usepackage{dcolumn}
\usepackage{ragged2e}
\newtheorem{thm}{Theorem}
\usepackage{hyperref}
\hypersetup{
    colorlinks=true,
    linkcolor=red,
    citecolor=blue,
    filecolor=magenta,      
    urlcolor=cyan,
    pdftitle={Can $f(R)$ gravity isotropize a pre-bounce contracting universe?},
    pdfpagemode=FullScreen} 
\usepackage{amsmath}
\usepackage{xcolor}
\usepackage{orcidlink}

\title{Can $f(R)$ gravity isotropise a pre-bounce contracting universe?}
\author[a]{Simran Arora\orcidlink{0000-0003-0326-8945}}
\author[a]{Sanjay Mandal\orcidlink{0000-0003-2570-2335}}
\author[b,a]{Saikat Chakraborty\orcidlink{0000-0002-5472-304X}}
\author[c,d]{Genly Leon\orcidlink{0000-0002-1152-6548}}
\author[a]{and P.K. Sahoo\orcidlink{0000-0003-2130-8832}}
\affiliation[a]{Department of Mathematics, Birla Institute of Technology and
Science-Pilani,\\ Hyderabad Campus, Hyderabad-500078, India} 
\affiliation[b]{Center for Space Research, North-West University, Mahikeng 2745, South Africa}
\affiliation[c]{Departamento de Matem\'{a}ticas, Universidad Cat\'{o}lica del Norte, Avda. Angamos 0610, Casilla 1280 Antofagasta, Chile} 
\affiliation[d]{Institute of Systems Science, Durban University of Technology, PO Box 1334, Durban 4000, South Africa} 
\emailAdd{dawrasimran27@gmail.com}
\emailAdd{sanjaymandal960@gmail.com}
\emailAdd{saikatnilch@gmail.com}
\emailAdd{genly.leon@ucn.cl}
\emailAdd{pksahoo@hyderabad.bits-pilani.ac.in}
\abstract{We address the important issue of isotropisation of a pre-bounce contracting phase in $f(R)$ gravity, which would be relevant to constructing any viable nonsingular bouncing scenario in $f(R)$ gravity. The main motivation behind this work is to investigate whether the $f(R)$ gravity, by itself, can isotropise a contracting universe starting initially with small anisotropy without incorporating a super-stiff or non-ideal fluid, impossible in general relativity. Considering Bianchi I cosmology and employing a dynamical system analysis, we see that this is not possible for $R^n$ ($n>1$) and $R+\alpha R^2$ ($\alpha>0$) theory, but possible for $\frac{1}{\alpha}e^{\alpha R}$ ($\alpha>0$) theory. On the other hand, if one does not specify an $f(R)$ theory a priori but demands a cosmology smoothly connecting an ekpyrotic contraction phase to a nonsingular bounce, the ekpyrotic phase may not fulfil the condition for isotropisation and physically viability simultaneously.}
\date{\today}
\keywords{Modified gravity, f(R) gravity, ekpyrotic contraction phase, nonsingular bounce}
\arxivnumber{2207.08479}
\begin{document}
\maketitle

\section{Introduction}
The issue related to the stability of homogeneous and isotropic cosmological solutions for small anisotropic perturbations has been studied intensely in theoretical cosmology \cite{Wainwright:1998ms, Chen:2001fh, Chen:2002ksa, Barrow:2000ka}. Behavior of small anisotropy has been studied in cosmological models, using  general relativity (GR), in the contexts of inflation \cite{Pereira:2015pga,Anninos:1991ma,Kitada:1992uh,Kitada:1992uf,Do:2017qyd} and pre-bounce ekpyrotic contraction phase \cite{Garfinkle:2008ei,Bozza:2009jx,Barrow:2010rx,Barrow:2015wfa}.  Generically, in an expanding anisotropic universe, any initially existing small anisotropy dies away as $\sim\frac{1}{a^6}$ ($a$ being the averaged scale factor). There are some exceptions. In the context of inflation in GR, the `No-Hair theorem' by Wald \cite{Wald:1983ky} asserts that all the Bianchi models except Bianchi IX isotropise during inflation (Bianchi IX model recollapses). In a pre-bounce ekpyrotic contracting phase $a(t)\sim(-t)^n$ ($t<0,\,0<n<1$), however, any initially existing small classical anisotropy grows as $\sim\frac{1}{a^6}$, faster (as $a\rightarrow 0$) than the energy density of any reasonable matter component, say, with energy density $\rho\sim\frac{1}{a^{3(1+\omega)}}$ with barotropic equation of state (e.o.s.) parameter $\omega=\frac{P}{\rho}$ satisfying $-1<\omega<1$. Therefore, even if one starts with a slightly perturbed FLRW universe, one ends up in a highly anisotropic Bianchi universe as the universe contracts. While working in GR, 	it is only provided that the universe is dominated by some matter component mimicking a super-stiff barotropic fluid with  e.o.s. parameter $\omega>1$, growth of small classical anisotropy can be suppressed for the growth of average Hubble parameter \cite{Erickson:2003zm, Barrow:2015wfa}. Without such fluid in a contracting phase, any initially existing small anisotropy grows large and dominates over all other matter components. That leads to the Belinsky-Khalatnikov-Lifshitz (BKL) instability \cite{Belinsky:1970ew}, either resulting in a post-bounce universe too anisotropic to be observationally viable or foiling a subsequent bounce altogether. Some works about bouncing universe in $f(R)$ and in other modified theories and their relation to the BKL instability connected with the anisotropy issue in such bounce scenario were investigated in \cite{Odintsov:2020zct, Nojiri:2022xdo, Odintsov:2021yva, Elizalde:2020zcb}. For proof of the no bounce behaviour in a scalar field dominated Bianchi I, Bianchi III and Kantowski-Sachs universe, see Ref. \cite{Solomons:2001ef}. Such a super-stiff fluid is usually mathematically realized by a fast-rolling scalar field \cite{Sharma:2015hka,Panda:2015wya,Solomons:2001ef,Cai:2013kja,Cai:2014bea}. Although the inclusion of a super-stiff matter component is by far the most popular choice for anisotropy suppression in a contracting universe, a choice for the same purpose of relaxing the requirement $w>1$ is non-ideal fluids, i.e. fluids that have in its energy-momentum tensor terms nonlinear in $\rho$ \cite{Dunsby:2003sr, Bozza:2009jx}. A super stiff matter component or a non-ideal fluid becomes a necessary but ad-hoc inclusion in the model.

The present paper aims to investigate whether the issue of such an ad-hoc inclusion can be alleviated with modified gravity theories. Modifications to Einstein's GR at the classical level become relevant at the high curvature regime, which includes the ekpyrotic contraction phase as it approaches the big crunch. Some modified  theories of gravity that arise by extending the Einstein-Hilbert action in a suitable way are Lovelock theories \cite{Lovelock:1971yv,Deruelle:1989fj},  Horndeski gravity 
\cite{Horndeski:1974wa},   generalized galileon theories  \cite{DeFelice:2010nf,Deffayet:2011gz,Dimakis:2017kwx,DeArcia:2015ztd,DeArcia:2015ztd}, equivalent  torsional
formulation of gravity \cite{Pereira,Xu:2012jf,Maluf:2013gaa,MKrvvak2019}, $f(T)$ gravity \cite{Cai:2015emx,Ferraro:2006jd,Linder:2010py}, $f(T,T_{G})$ gravity
\cite{Kofinas:2014owa,Kofinas:2014aka}, $f(T,B)$ gravity \cite{Bahamonde:2015zma,Paliathanasis:2021ysb,Paliathanasis:2022pgu}, etc. Scalar fields can be introduced to construct 
scalar-torsion theories \cite{Cid:2017wtf,Leon:2022oyy,Geng:2011aj}. These allow for non-minimal \cite{Geng:2011aj,Geng:2011ka,Gonzalez-Espinoza:2020jss,Paliathanasis:2021nqa,Gonzalez-Espinoza:2021qnv,Toporensky:2021poc} or derivative  
\cite{Kofinas:2015hla,Kofinas:2014aka} couplings with torsion, or more general constructions  
\cite{Geng:2012vtr,Skugoreva:2014ena,Jarv:2015odu,Skugoreva:2016bck,Hohmann:2018rwf,Hohmann:2018vle,Hohmann:2018ijr,Hohmann:2018dqh,Emtsova:2019qsl}, including the teleparallel version of Horndeski theories \cite{Bahamonde:2019shr,Bahamonde:2020cfv,Bahamonde:2021dqn,Bernardo:2021izq}. This paper will consider one of the simplest modified gravity theories, namely $f(R)$ gravity, which belongs to the Horndeski classification. $f(R)$ gravity is a straightforward generalisation of Einstein's gravity obtained by replacing the Ricci scalar $R$ in the Einstein-Hilbert action with a smooth function of $R$: $R\to f(R)$. Different stability conditions require $f'(R), f''(R)>0$ \cite{Sotiriou:2008rp,DeFelice:2010aj,Nojiri:2005jg,DeFelice:2008wz,Leon:2014yua,Leon:2013bra,Leon:2011zpk,Leon:2010pu}. 

We will consider the metric formulation of $f(R)$ gravity, in which the metric is assumed to be the only dynamical degree of freedom to which variation of the action is to be taken. A different formulation of $f(R)$ gravity is called the Palatini formulation, which considers both the metric and the connection to be independent dynamical degrees of freedom for which variation of the action is to be taken. While these two formulations give identical field equations in GR, in the case of $f(R)$ gravity, they give different field equations \footnote{See Ref. \cite{Capozziello:2010ef} for a correspondence between the two formalisms in $f(R)$ gravity.}. 

The stability of an isotropic ekpyrotic contraction phase for small anisotropy, or the evolution of metric anisotropy during a contraction phase in general, has not yet been explored much in the context of modified gravity theories. In Ref. \cite{Barragan:2010qb}, the authors explicitly showed that it is possible to get a bouncing Bianchi I cosmology in a modified gravity of the form $f(R, R_{\mu\nu})=R+aR^{2}+bR_{\mu\nu}R^{\mu\nu}$ in Palatini formalism where the metric anisotropy remains bounded. In Ref. \cite{Bhattacharya:2017cbn}, working with Bianchi I cosmology and $R+\alpha R^2$ gravity with $\alpha<0$, the authors attempted an exact solution for the evolution of metric anisotropy in some particular simple cases. In particular, the authors could show that during an ekpyrotic contraction phase $a(t)\sim(-t)^n$ ($t<0,\,0<n<1$), depending on the initial anisotropy, there was a range for the parameter $n$, for which anisotropy could decrease along the contraction. A different approach was taken in Ref. \cite{Chakraborty:2018thg}, in which the author tried to generalise the well-known reconstruction method of $f(R)$ gravity to anisotropic Bianchi I spacetimes. Demanding that metric anisotropy should decrease exponentially fast for the averaged isotropic evolution during an ekpyrotic contraction, the author could, in principle, reconstruct the $f(R)$ model that can achieve this. However, the resulting $f(R)$ model did not come in any compact form and was too obscure for further analysis.

In this paper, we address the issue of the stability of an isotropic ekpyrotic contraction phase for small anisotropy with the help of the dynamical system approach, which, we believe, gives a clearer picture than the relevant earlier attempts. We confine ourselves to Bianchi I cosmology, which, as shown in the next section, can be interpreted as the homogeneously perturbed spatially flat FLRW universe as long as the anisotropy is small. We follow the dynamical system formulation for Bianchi I cosmology in $f(R)$ gravity as was given in Refs. \cite{Chakraborty:2018bxh,Chakraborty:2018ost}. We do the analysis for some specific simple $f(R)$ models as well as carry out a \emph{form-independent} analysis using kinematic parameters \cite{Chakraborty:2021mcf}.

This research is divided into two parts. First, we seek particular cases when the metric anisotropy does not become dominating during a contracting phase, establishing some general conditions. As we will see, only specific $f(R)$ theories may allow for such behaviour, and we identify some of them. However, even though an isotropising contracting solution is found, there is no guarantee that this will lead to a subsequent bounce. That might not be a big issue as it is common in constructing nonsingular bouncing paradigms to incorporate two scalar d.o.f. \cite{Cai:2013kja}, one responsible for the isotropisation and one responsible for the bounce. Nonetheless, it would be interesting if we could achieve both via one single scalar d.o.f. The bounce is postulated in the second part, where we deal with the anisotropy behaviour for an ekpyrotic contraction phase smoothly connecting to a nonsingular bounce.

The paper is organised as follows. In Sec. \ref{sec:hom_pert} we state our motivation for working with Bianchi I spacetime by showing how it can be interpreted as the homogeneously perturbed FLRW spacetime in the synchronous gauge. In Sec. \ref{sec:bianchi1_GR} we review the anisotropy problem in a pre-bounce contracting universe in GR and the need for a super-stiff fluid. In Sec. \ref{sec:bianchi1_f(R)} we write down the field equations for Bianchi I spacetime in $f(R)$ gravity. In Sec. \ref{sec:dynsys} we present a dynamical system formulation for Bianchi I cosmology in $f(R)$ gravity. In Sec. \ref{sec:examples} we apply the dynamical system formulation to three specific simple but popular $f(R)$ models. In Sec. \ref{sec:form-independent} we carry out a \emph{form-independent} dynamical system analysis by considering an ansatz cosmological evolution that represents a smooth transition from an ekpyrotic contraction phase to a nonsingular bounce. We discuss our results and conclude in Sec. \ref{sec:ending}. We also give some necessary technical details in Appendices \ref{app:jacobian_eigenvalues},  \ref{app:compactification} and \ref{AppC}, which may be helpful for an interested reader.  
\section{Bianchi I cosmology}\label{sec:hom_pert}
Consider the spatially flat case of the homogeneous and isotropic FLRW cosmology
\begin{equation}\label{flrw}
    ds^2 = - dt^2 + a^{2}(t)\delta_{ij}dx^{i}dx^{j}.
\end{equation}
Small scale inhomogeneity and anisotropy on top of an otherwise homogeneous and isotropic cosmological background can be treated as cosmological perturbations. Perturbed spatially flat FLRW metric can be written as
\begin{equation}
  ds^2 = - (1+2\phi(\bar{x},t))dt^2 - 2B_{i}(\bar{x},t)dtdx^i + a^{2}(t)[(1-2\psi(\bar{x},t))\delta_{ij} + 2h_{ij}(\bar{x},t)]dx^{i}dx^{j}, 
\end{equation}
where $\phi$, $\psi$, $B_i$ and $h_{ij}$ are spacetime dependent perturbation quantities. The above $\psi$ is the trace part, and $h_{ij}$ is the traceless part of $\delta g_{ij}$. In Synchronous gauge ($\phi=0,\,B_i=0$) the perturbed metric becomes
\begin{small}
\begin{equation}
    ds^2 = -dt^2 + a^{2}(t)[(1-2\psi(\bar{x},t))\delta_{ij} + 2h_{ij}(\bar{x},t)]dx^{i}dx^{j}.
\end{equation}
\end{small}
Since we are only concerned about small anisotropy in this work, we assume that any inhomogeneity is negligible. In other words, we only consider homogeneous perturbations
\begin{equation}
    ds^2 = -dt^2 + a^{2}(t)[(1-2\psi(t))\delta_{ij} + 2h_{ij}(t)]dx^{i}dx^{j}.
\end{equation}
Suppose that $\beta_{1}(t),\,\beta_{2}(t),\,\beta_{3}(t)$ are the eigenvalues of the $3\times3$ matrix $h_{ij}(t)$. Since $h_{ij}$ is traceless, $\beta_{1}+\beta_{2}+\beta_{3}=0$. If one changes the basis ${x^i}$ to the eigenvector basis of the matrix $h_{ij}(t)$, one can write the spatial part of the perturbed metric as
\begin{small}
\begin{equation}
    g_{ij}(t) = a^{2}(t)\left[(1-2\psi(t))\delta_{ij}+diag(2\beta_{1}(t),2\beta_{2}(t),2\beta_{3}(t))\right],
\end{equation}
\end{small}
or, since $\psi$ and $\beta_i$s are perturbation quantities, as,
\begin{small}
\begin{equation}
    g_{ij}(t) = a^{2}(t) \text{diag}\left(e^{-2\psi(t)+2\beta_{1}(t)},e^{-2\psi(t)+2\beta_{2}(t)},e^{-2\psi(t)+2\beta_{3}(t)}\right).
\end{equation}
\end{small}
One can absorb the trace part $e^{-2\psi(t)}$ into $a(t)$. What one arrives at is a parametrised form of the spatially flat homogeneous and anisotropic Bianchi I metric
\begin{small}
\begin{equation}\label{bianchi-1}
    ds^2 = - dt^2 + a^{2}(t)[e^{2\beta_1(t)}dx_1^2 + e^{2\beta_2(t)}dx_2^2 + e^{2\beta_3(t)}dx_3^2].
\end{equation}
\end{small}
In general, the homogeneous and anisotropic Bianchi I metric is one in which all the three scale factors in three orthogonal spatial directions are different
\begin{equation}
    ds^2 = - dt^2 + a_1^{2}(t)dx_1^2 + a_2^{2}(t)dx_2^2 + a_3^{2}(t)dx_3^2.
\end{equation}
The corresponding Hubble parameters are defined as $H_i=\frac{\dot{a}_i}{a_i}$ for $i=1,2,3$. One can define an average scale factor as the geometric mean of the three scale factors
\begin{equation}
a(t)=[a_1(t)a_2(t)a_3(t)]^{1/3}.
\end{equation}
The average Hubble parameter $H(t)$ derived from the average scale factor is, therefore, the arithmetic mean of the three Hubble parameters
\begin{equation}
H(t)=\frac{1}{3}[H_1(t)+H_2(t)+H_3(t)].
\end{equation} 
If one uses the parametrisation 
\begin{equation}
a_i(t)=a(t)e^{\beta_i(t)}, \qquad H_i(t)=H(t)+\dot{\beta}_i(t),
\end{equation}
with the three parameters $\beta_1(t),\beta_2(t),\beta_3(t)$ constrained by the relation $\beta_1(t)+\beta_2(t)+\beta_3(t)=0$, then one arrives at the parametrized form in Eq. \eqref{bianchi-1}. Observe that when $\dot{\beta}_1=\dot{\beta}_2=\dot{\beta}_3=0$, one can suitably rescale the spatial coordinates to get back the FLRW metric. One can therefore define a quantity $\sigma$ as 
\begin{equation}\label{sigma}
    \sigma^2 = \dot{\beta}_1^2 + \dot{\beta}_2^2 + \dot{\beta}_3^2,
\end{equation} 
such that $\sigma=0$ implies $\dot{\beta}_1=\dot{\beta}_2=\dot{\beta}_3=0$, i.e. the universe is isotropic. Notice that $\sigma^2 \propto \sigma^{ij}\sigma_{ij}$, where $\sigma_{ij}$ is the shear tensor defined as
\begin{equation}
    \sigma_{ij}=\frac{1}{2}\dot{\gamma}_{ij},
\end{equation}
$\gamma_{ij}=(e^{2\beta_i})\delta_{ij}$ being the 3-metric on the constant time hypersurface.
\section{Bianchi I cosmology in GR and need for an ekpyrotic contraction phase}\label{sec:bianchi1_GR}
In this section, we write the cosmological field equations for Bianchi I spacetime in GR and explain the need for an ekpyrotic contraction phase before the bounce while modelling a bouncing cosmology in GR. To explain the issue of anisotropy divergence during a contracting phase, it suffices to consider only the case of an isotropic fluid. Therefore we confine ourselves to this case in this section.

The Einstein equations and the continuity equation in the presence of an isotropic perfect fluid given by an energy-momentum tensor
\begin{equation}
    T^{\mu}_{\nu} = \text{diag}(-\rho,P,P,P),
\end{equation}
for a Bianchi I spacetime are as follows:
\begin{subequations}
\begin{eqnarray}
&& 3H^2 = \kappa\rho + \frac{1}{2}\sigma^2, \label{bianchi_gr_1}\\
&& 2\dot{H} + 3H^2 = - \kappa P - \frac{1}{2}\sigma^2, \label{bianchi_gr_2}\\
&& \dot{\sigma} + 3H\sigma = 0, \label{bianchi_gr_3}\\ && \dot{\rho} + 3H\rho(1+w) = 0, \label{cont_pf}
\end{eqnarray}
\end{subequations}
where $w=\frac{P}{\rho}$ is the barotropic equation of the state parameter of the fluid. From Eq. \eqref{bianchi_gr_3} one can see that $\sigma^2 \sim \frac{1}{a^6}$. On the other hand, for a barotropic fluid ($w=$constant) energy density of a perfect fluid goes as $\rho\sim\frac{1}{a^{3(1+\omega)}}$. During an inflationary epoch, the universe rapidly expands, so any pre-existing small metric anisotropy diminishes faster than a matter component with e.o.s parameter in the range $-1<\omega<1$ (which includes nonrelativistic and relativistic fluids), and the universe quickly isotropises. For a massless scalar field, however, the effective equation of state is $\omega_{\phi}=P_{\phi}/\rho_\phi=1$. Therefore the energy density of a massless scalar field decays at the same rate as the metric anisotropy $\rho_\phi \sim \sigma \propto a^{-6}$. However, in a pre-bounce contracting epoch, the anisotropic contribution to the right-hand side of the Friedmann constraint \eqref{bianchi_gr_1} goes up faster than any matter component with the e.o.s parameter in the range $-1<\omega<1$. That leads to a highly anisotropic vacuum universe given by a Kasner solution \cite{Erickson:2003zm} as the universe approaches the big crunch, even if the initial anisotropy was perturbatively small. 

As discussed in Ref. \cite{Erickson:2003zm}, one way to suppress anisotropy during a contracting universe is to invoke a super-stiff ($\omega>1$) matter component. That is usually achieved by a scalar field rolling down a steep negative potential \cite{Xue:2011nw, Xue:2013iqy}. To illustrate this with a heuristic example, consider an FLRW universe dominated by a scalar field with the potential
\begin{equation}
    V(\phi) = - V_0 e^{-\sqrt{\frac{2}{\alpha}}\phi}, 
\end{equation}
where $0<\alpha\ll1$ and $V_0>0$. The equations of motion
\begin{subequations}
\begin{eqnarray}
&& 3H^2 = \frac{1}{2}\dot{\phi}^2 + V(\phi), \\
&& \dot{H} = -\frac{1}{2}\dot{\phi}^2, \\
&& \ddot{\phi} + 3H\dot{\phi} + \frac{dV}{d\phi} = 0,
\end{eqnarray}
\end{subequations}
admit the attractor solution
\begin{subequations}
\begin{eqnarray}
&& a\sim(-t)^\alpha, \label{ekpyrotic}\\
&& H = \frac{\alpha}{t},\\
&& \phi(t) = \sqrt{2\alpha}\ln\left[-\sqrt{\frac{V_0}{\alpha(1-3\alpha)}}\,\,t\right],
\end{eqnarray}
\end{subequations}
where $t$ is negative and increases towards zero. The equation of state parameter for this attractor solution is
\begin{equation}
    \omega_\phi = - 1 + \frac{2}{3\alpha}
\end{equation}
which is $>1$ as long as $0<\alpha<\frac{1}{3}$. The slow contraction phase, as given by the time evolution in Eq. \eqref{ekpyrotic} is called an ekpyrotic phase. Apart from solving the anisotropy problem during contraction, such a phase is also in line with the requirements to solve the horizon problem in the nonsingular bouncing paradigm.
\section{Bianchi I cosmology in $f(R)$ gravity}\label{sec:bianchi1_f(R)}
The evolution of metric anisotropy in $f(R)$ gravity is not straightforward. This section builds the framework for studying the Bianchi I cosmology in $f(R)$ gravity following Refs. \cite{Chakraborty:2018bxh, Chakraborty:2018ost}, which we will use in the subsequent dynamical system analysis. For completeness, we will also consider the case of an anisotropic fluid here.

Following \cite{Chakraborty:2018bxh,Chakraborty:2018ost} we choose to work with the parameters 
\begin{equation}\label{beta_pm}
\beta_{\pm}=\beta_1\pm\beta_2,
\end{equation}
so that the quantity $\sigma^2$ from Eq. \eqref{sigma} can be written as
\begin{equation}
\sigma^2 = \dot{\beta}_1^2 + \dot{\beta}_2^2 + \dot{\beta}_3^2 = \frac{3}{2}\dot{\beta}_+^2 + \frac{1}{2}\dot{\beta}_-^2.
\end{equation}
The energy-momentum tensor for the anisotropic perfect fluid reads
\begin{equation}
T^{\nu}_{\mu} = \text{diag}(-\rho,P_1,P_2,P_3)=\text{diag}(-\rho, \omega_1 \rho,\omega_2 \rho,\omega_3 \rho).
\end{equation}
One can define an average equation of state parameter $\omega$ and the deviations $\mu_i$ from the average equation of state as follows
\begin{equation}
    \omega = \frac{1}{3}(\omega_1 + \omega_2 + \omega_3), \quad \mu_i = \omega_i - \omega, \quad \omega_i = \omega + \mu_i,
\end{equation}
for $i=1,2,3$. Clearly, $\mu_i$s follow a similar constraint equation as the $\beta_i$s: $\mu_1 +\mu_2+\mu_3=0$. Similar to Eq. \eqref{beta_pm}, we choose to work with the parameters
\begin{equation}
  \mu_{\pm} = \mu_1 \pm \mu_2.
\end{equation}

Modified Einstein equations and the continuity equation are as follows:
\begin{subequations}
\begin{eqnarray}
&& 3H^2 = \frac{\kappa}{f'}\bigg(\rho+\frac{Rf'-f}{2\kappa} - \frac{3Hf''\dot{R}}{\kappa}\bigg) + \frac{\sigma^2}{2}, \label{bianchi_fr_1}\\
&& 2\dot{H} + 3H^2 + \frac{\sigma^2}{2} = - \frac{\kappa}{f'}\left[\omega\rho - \frac{Rf'-f}{2\kappa} + \frac{\dot{R}^{2}f''' + (2H\dot{R}+\ddot{R})f''}{\kappa}\right], \label{bianchi_fr_2}\\
&& \ddot{\beta}_\pm + \left(3H + \frac{\dot{R} f''}{f'}\right) \dot{\beta_{\pm}} = \frac{\kappa\rho}{f'}\mu_{\pm}, \label{bianchi_fr_31}\\
&& \dot{\rho} + \left[3H(1+\omega) + \frac{3}{2}\mu_{+}\dot{\beta_{+}} + \frac{1}{2}\mu_{-}\dot{\beta_{-}}\right]\rho = 0, \label{cont_pf_1}
\end{eqnarray}
\end{subequations}
For an isotropic perfect fluid, $\mu_+=\mu_-=0$, so that the anisotropy evolution equation \eqref{bianchi_fr_31} simplifies to
\begin{equation}\label{bianchi_fr_32}
\dot{\sigma} + \left(3H + \frac{\dot{R}f''}{f'}\right)\sigma=0
\end{equation}
whereas the continuity equation \eqref{cont_pf_1} simplifies back to Eq. \eqref{cont_pf}. In this case, the anisotropy equation can be solved to obtain
\begin{equation}\label{bianchi_fr_33}
    \sigma \sim \frac{1}{a^3 f'(R)}.
\end{equation}

One can see the problem one encounters when one tries to calculate for anisotropy evolution in $f(R)$ gravity from Eq. \eqref{bianchi_fr_33}. Given an $f(R)$, solving for $\sigma(a)$ requires the knowledge of $R(a)$. But, by definition, the Ricci scalar for the Bianchi I metric \eqref{bianchi-1} is
\begin{equation}
R = 6\dot{H} + 12H^2 + \sigma^2,
\end{equation}
which already contains the term $\sigma^2$. Therefore, even if an $f(R)$ is supplied, an analytic solution for $\sigma(a)$ is not to be found unless for some particular forms of $f(R)$ (see Ref. \cite{Bhattacharya:2017cbn}). the best approach to understanding the overall dynamics is to resort to dynamical system analysis, which we follow from the next section onwards. However, this very issue is also at the heart of the motivation behind this whole study. From Eq. \eqref{bianchi_fr_33}, one cannot naively say $\sigma^2 \sim\frac{1}{a^6}$. That gives them hope that maybe for some forms of $f(R)$, it might be possible to isotropise the universe during an ekpyrotic contraction without invoking any ad-hoc super-stiff matter by hand component.

\section{Dynamical system formulation for Bianchi I cosmology in $f(R)$ gravity}\label{sec:dynsys}

This section presents the dynamical system formulation for the homogeneous anisotropic Bianchi I cosmology in $f(R)$ gravity following Refs. \cite{Chakraborty:2018bxh,Chakraborty:2018ost}. We consider both isotropic and anisotropic cases as well as the vacuum case. For the matter case, we confine ourselves to barotropic matters, i.e. $\frac{P}{\rho}=w=$constant. The first step is to define a set of Hubble-normalized dimensionless dynamical variables \footnote{The reader is advised to note the difference in the definitions of $u_4^{\pm}$ with that in Refs.  {\cite{Chakraborty:2018bxh,Chakraborty:2018ost}}.}
\begin{align}
u_1  & =\frac{\dot{R}f''}{f'H},\,\, u_2 = \frac{R}{6H^2},\,\, u_3 = \frac{f}{6f'H^2}, u_4^+ = \frac{\dot{\beta}_+}{2H},\,\, u_4^- = \frac{\dot{\beta}_-}{2 \sqrt{3}H}, \nonumber\\
u_4^2 & = \left(u_4^+\right)^2 + \left(u_4^-\right)^2 = \frac{\sigma^2}{6H^2},\,\, u_5 = \frac{k \rho}{3f'H^2}.
\label{dynvars}
\end{align}
The next step is to define a phase spacetime variable. Since we will exclusively concentrate on the contracting phase for our purpose, we introduce the following phase spacetime variable
\begin{equation}
\label{time-tau}
\tau=-\ln a.
\end{equation}
This definition is justified as $\tau$ is a monotonically increasing function of $t$ for a contracting universe: $\dot{\tau}=-H>0$. It is important to characterise what is the ``past'' ($\tau \rightarrow -\infty$) or ``future''  ($\tau \rightarrow +\infty$) attractor in terms of this variable. Using the logarithmic time defined in Eq. \eqref{time-tau}, $\tau \rightarrow -\infty$ means  $\ln (a) \rightarrow +\infty \implies a\rightarrow + \infty$ and  $\tau \rightarrow +\infty$ means  $\ln (a) \rightarrow -\infty \implies a \rightarrow 0^+$. That is, in a contracting universe the past attractor corresponds to $a\rightarrow \infty $, and future attractor corresponds to $a\rightarrow  0^+$.  \footnote{Note that, in dealing with expanding cosmologies, we could use as the phase spacetime variable the e-folding number $N\equiv\ln a$, in dealing with contracting cosmologies, we had to define a different time-variable $\tau=-\ln a$. In an expanding universe the past attractor corresponds to $N\rightarrow -\infty \implies a\rightarrow 0^+$, and future attractor corresponds to $N\rightarrow +\infty \implies a\rightarrow +\infty$. While defining a ``time'' variable in the phase-space formulation, one must remember that the temporal variable must be a monotonically increasing function.} 

Lastly, as in the case of the dynamical system for isotropic cosmologies, we also introduce the auxiliary quantity 
\begin{equation}
\Gamma(R)\equiv\frac{f'}{Rf''}.
\end{equation}
Below we write the dynamical systems.
\begin{itemize}

\item Anisotropic fluid:   
For the generic case of anisotropic fluid, the Friedmann constraint equation is
\begin{equation}\label{fried_const}
1 + u_1 - u_2 + u_3 - u_4^2 - u_5 = 0.
\end{equation}
Choosing the constraint to eliminate $u_5$, the dynamical system can be written as
\begin{subequations}\label{dynsys_aniso}
\begin{align}
& \frac{du_1}{d\tau} = u_1^2 + u_1 \left(u_2 - u_4^2\right) + 3\omega\left(u_1 - u_2 + u_3 - u_4^2 + 1\right) - u_2 + 3u_3 + u_4^2 - 1,\\
& \frac{du_2}{d\tau} = 2u_2 \left(u_2 - u_4^2 - 2\right) - u_1 u_2 \Gamma,\\
& \frac{du_3}{d\tau} = u_3 \left(u_1 + 2u_2 - 2u_4^2 - 4\right) - u_1 u_2 \Gamma,\\
& \frac{du_4^+}{d\tau} = - \frac{3}{2}\mu_+ \left(u_1 - u_2 + u_3 - u_4^2 + 1\right) + u_4^+ \left(u_1 + u_2 - u_4^2 + 1\right),\\
& \frac{du_{4}^-}{d\tau} = - \frac{\sqrt{3}}{2}\mu_- \left(u_1 - u_2 + u_3 - u_4^2 + 1\right) + u_4^- \left(u_1 + u_2 - u_4^2 + 1\right).
\end{align}
\end{subequations}
\item Isotropic fluid: For isotropic fluid ($\mu_+ = \mu_- = 0$) the Friedmann constraint remains the same as in Eq. \eqref{fried_const}, but the dynamical system \eqref{dynsys_aniso} can be simplified to 
\begin{subequations}\label{dynsys_iso}
\begin{align}
& \frac{du_1}{d\tau} = u_1^2 + u_1 \left(u_2 - u_4^2\right) + 3\omega\left(u_1 - u_2 + u_3 - u_4^2 + 1\right) - u_2 + 3u_3 + u_4^2 - 1,\\
& \frac{du_2}{d\tau} = 2u_2 \left(u_2 - u_4^2 - 2\right) - u_1 u_2 \Gamma,\\
& \frac{du_3}{d\tau} = u_3 \left(u_1 + 2u_2 - 2u_4^2 - 4\right) - u_1 u_2 \Gamma,\\
& \frac{du_{4}}{d\tau} = u_4 \left(u_1 + u_2 - u_4^2 + 1\right). \label{eq:34d}
\end{align}
\end{subequations}

\item Vacuum: for vacuum ($u_5=0$) the Friedmann contraint \eqref{fried_const} simplifies to
\begin{equation}\label{fried_const_vac}
    1 + u_1 - u_2 + u_3 - u_4^2 = 0.
\end{equation}
Choosing to eliminate $u_3$ using the constraint, the dynamical system \eqref{dynsys_iso} can be simplified to 
\begin{subequations}\label{dynsys_vac}
\begin{align}
& \frac{du_1}{d\tau} = u_1^2 + u_1 \left(u_2 - u_4^2 - 3\right) + 2\left(u_2 + 2u_4^2 - 2\right),\\
& \frac{du_2}{d\tau} = - u_1 u_2 \Gamma + 2u_2 \left(u_2 - u_4^2 - 2\right),\\
& \frac{du_4}{d\tau} = u_4 \left(u_1 + u_2 - u_4^2 + 1\right). 
\end{align}
\end{subequations}
\end{itemize}
As is well known, the success of this particular dynamical system formulation depends crucially on the invertibility of the relation
\begin{equation}
\frac{u_2}{u_3} = \frac{Rf'}{f}    
\end{equation}
to obtain $R=R\left(\frac{u_2}{u_3}\right)$, because only then $\Gamma(R)$ can be expressed as a function of $u_2$ and $u_3$, $\Gamma(R) \equiv h\left(u_2,u_3\right)$ and the dynamical systems \eqref{dynsys_aniso} or \eqref{dynsys_iso} can be closed. In case of vacuum, we have chosen to eliminate $u_3$ using the constraint \eqref{fried_const_vac}, so that now $\Gamma(R)$ needs to be expressed as a function of $u_1,\,u_2,\,u_4$: $\Gamma(R) \equiv l\left(u_1,u_2,u_4\right)$.

Before moving on to the next section, we mention that, as one can note from the dynamical systems \eqref{dynsys_iso} and \eqref{dynsys_vac}, $u_4=0$ is an invariant submanifold, which contains all the isotropic cosmologies. That is not the case for the system \eqref{dynsys_aniso}. That intuitively makes sense because, in the presence of an anisotropic fluid, the anisotropy of the fluid (as characterised by nonzero $\mu_\pm$) will induce anisotropy in the spacetime geometry. There is no such anisotropy induction in the case of an isotropic fluid or the absence of any fluid.

When metric anisotropy is small and can be considered as homogeneous perturbations on top of an otherwise isotropic cosmological evolution, as in the treatment of Sec. \ref{sec:hom_pert}, one can find conditions for ``isotropisation'', based on the stability of the isotropic manifold $u_4=0$ under anisotropic perturbations. 
When $u_4\neq 0$, there is an ``exact'' stability condition for the isotropic manifold, that  is, \emph{if $u_4\neq 0$, the anisotropic perturbations decay as far 
\begin{equation}\label{exact-isotropisation}
    \left(u_1 + u_2   + 1\right)<u_4^2.
\end{equation}}
When the terms of order $u_4^2$ are neglected, we obtain the weak condition 
\begin{equation}\label{isotropisation}
 u_1 + u_2 + 1 < 0.
\end{equation}
These conditions are found from the $u_4$-equation(s), by demanding that the submanifold $u_4=0$ is attracting for initial values $u_4\neq 0$: $\left[\frac{\partial}{\partial u_4}\frac{du_4}{d\tau}\right]_{u_4 \rightarrow 0} < 0$ for isotropic fluid and vacuum, $\left[\frac{\partial}{\partial u_4^{\pm}}\frac{du_4}{d\tau}\right]_{u_4^\pm \rightarrow 0} < 0$ for anisotropic fluid. If an isotropic equilibrium point falls on this region, one can say that any small positive perturbative anisotropy will die out in the cosmic phase represented by that equilibrium point. However, if $u_1 + u_2 + 1 = 0$ ($u_4\neq 0$), then from equation \eqref{eq:34d} it follows 
\begin{equation}
    \frac{du_4}{d\tau} = -u_4^3 =-\nabla U(u_4), \quad  U(u_4)= \frac{1}{4} u_4^4.  
\end{equation}
That is a gradient-like equation under a quartic potential. Then, $u_4=0$ is a degenerated minimum of a second order of the quartic potential, which means it is stable. Hence, at first order in the anisotropies (``small anisotropy''), the stability region of isotropisation of the isotropic invariant submanifold $u_4 =0$  is generalised to
\begin{equation}\label{isotropisation}
 u_1 + u_2  + 1 \leq 0.
\end{equation}

\section{Specific examples}\label{sec:examples}

In this section we consider some specific example $f(R)$ forms. Since our goal here is to investigate the isotropisation properties of these $f(R)$ forms during contraction, we will not carry out an extensive phase space analysis. Rather, we will look for isotropic equilibrium points and investigate their stability properties. Moreover, one can note that, for the case of an anisotropic fluid, as apparent from the $u_{4\pm}$-equation in the system \eqref{dynsys_aniso}, the existence of isotropic equilibrium points ($u_4=0$) demands the condition
\begin{equation}
1 + u_1 - u_2 + u_3 = 0 = u_5,   
\end{equation}
to be satisfied at the equilibrium points, i.e. the corresponding cosmologies must also be a vacuum. That will, therefore, give the same isotropic equilibrium points as what can be obtained from the dynamical system \eqref{dynsys_vac} for the vacuum. The Jacobian, and hence the nature of stability, however, will generally differ in the two cases.

We consider three examples below
\begin{itemize}
    \item $f(R) = R^n$ with $n>1$.
    \item $f(R) = R + \alpha R^2$ with $\alpha>0$.
    \item $f(R) = \frac{1}{\alpha}e^{\alpha R}$ with $\alpha>0$.
\end{itemize}
For each example, we consider the cases of anisotropic fluid, isotropic fluid and no fluid. In respective tables, we list the isotropic equilibrium points and their stable nature as found by a Jacobian analysis. Jacobian eigenvalues are given in the appendix for the reader's reference. The conditions for the absence of ghost and tachyonic instability ($f'>0,\,f''>0$) and the behaviour of small perturbative anisotropy as specified by the condition \eqref{isotropisation} will be presented as possible for each equilibrium point.
\subsection{First example: $f(R) = R^{n}$, $n>1$}
As a first example, let us consider the monomial gravity model $f(R) = R^n$ with $n>1$. This model can be interpreted as the high curvature limit of a more generic polynomial $f(R)$ model of degree $n$. Absence of ghost and tachyonic instability requires $f',\,f''>0$, which implies
\begin{equation}
    R^{n-1}>0, \quad R^{n-2}>0.
\end{equation}
To guarantee that both conditions are satisfied, $R$ must be a definite positive. For this particular model, the analysis becomes simple because one has
\begin{equation}
\frac{u_2}{u_3}=n,
\end{equation}
that provides an additional constraint to reduce the dimensionality of the phase space by one. For this particular $f(R)$ model $\Gamma$ is just a constant
\begin{equation}
    \Gamma = \frac{1}{n-1}.
\end{equation}
The isotropic equilibrium points in presence of anisotropic fluid, isotropic fluid and in absence of any fluid are listed in tables \ref{tab:fp_mon_aniso}, \ref{tab:fp_mon_iso} and \ref{tab:fp_mon_vac} respectively. 
\begin{table}[h]
\resizebox{\textwidth}{!}{
    \centering
    \begin{tabular}{|c|c|c|c|c|c|}
   \hline
  
   Point & \multicolumn{1}{|p{4cm}|}{\centering Coordinates \\ $(u_1,u_2,u_3,u_4^{\pm},u_5)$} & Stability & Cosmology & Sign of $R_*$ & \multicolumn{1}{|p{4cm}|}{\centering Behavior of small\\ anisotropy} \\
   \hline
   $\mathcal{P}_1$ & $\left(-1,0,0,0,0\right)$ & Non-hyperbolic  & $a\sim(-t)^{1/2}$  & $R_*=0$ & Decreasing \\
   \hline
    $\mathcal{P}_2$ &  $\scriptscriptstyle\left(\frac{2(n-2)}{2n-1},\frac{n(4n-5)}{(n-1)(2n-1)},\frac{(4n-5)}{(n-1)(2n-1)},0,0\right)$ & \multicolumn{1}{|p{4cm}|}{\centering Stable for $1<n<\frac{5}{4}$,  $3\omega<\frac{13n-3-8n^2}{(n-1)(2n-1)}$, \\ Unstable for $n>\frac{5}{4}$,  $3\omega>\frac{13n-3-8n^2}{(n-1)(2n-1)}$, \\ Non-hyperbolic or saddle otherwise} & $a\sim(-t)^{\frac{2n^2 - 3n +1}{2-n}}$  & $R_*\gtreqless0$ for $n\gtreqless\frac{5}{4}$ & \multicolumn{1}{|p{4cm}|}{\centering Decreasing for \\ $1<n\leq\frac{5}{4}$, \\ Increasing for \\ $n>\frac{5}{4}$.} \\
    \hline
\end{tabular}}
    \caption{Isotropic equilibrium points for $R^n$ gravity ($n>1$) in presence of an anisotropic fluid. The physical viability of the solution corresponding to an equilibrium point can be guaranteed when $R_* = 6H_*^{2}u_{2*}>0$, where the subscript $*$ implies values calculated at the equilibrium point. The behaviour of small anisotropy is investigated via Eq. \eqref{isotropisation}. The isotropic equilibrium point $\mathcal{P}_2$ can represent a stable contracting power-law cosmology with isotropisation when $1<n<\frac{5}{4}$ and $3\omega<\frac{13n-3-8n^2}{(n-1)(2n-1)}$, but physically viability of this solution cannot be guaranteed since $R_*<0$.}
    \label{tab:fp_mon_aniso}
\end{table}
\begin{table}[h]
\resizebox{\textwidth}{!}{
    \centering
    \begin{tabular}{|c|c|c|c|c|c|}
   \hline
   Point & \multicolumn{1}{|p{2cm}|}{\centering Coordinates \\ $(u_1,u_2,u_3,u_4,u_5)$} & Stability & Cosmology & Sign of $R_*$ & \multicolumn{1}{|p{3cm}|}{\centering Behavior of small\\ anisotropy} \\
   \hline
     $\mathcal{P}_1$ & $\left(-1,0,0,0,0\right)$ & Non-hyperbolic  & $a\sim(-t)^{1/2}$  & $R_*=0$ & Decreasing \\
   \hline 
   $\mathcal{P}_2$ & $\scriptscriptstyle\left(\frac{2(n-2)}{2n-1},\frac{n(4n-5)}{(n-1)(2n-1)},\frac{(4n-5)}{(n-1)(2n-1)},0,0\right)$ & \multicolumn{1}{|p{4cm}|}{\centering Stable for $1<n<\frac{5}{4}$,  $3\omega<\frac{13n-3-8n^2}{(n-1)(2n-1)}$, \\ Unstable for $n>\frac{5}{4}$,  $3\omega>\frac{13n-3-8n^2}{(n-1)(2n-1)}$, \\ Non-hyperbolic or saddle otherwise} & $a\sim(-t)^{\frac{2n^2 - 3n +1}{2-n}}$  & $R_*\gtreqless0$ for $n\gtreqless\frac{5}{4}$ & \multicolumn{1}{|p{3cm}|}{\centering Decreasing for \\ $1<n\leq\frac{5}{4}$, \\ Increasing for \\$n>\frac{5}{4}$.} \\
   \hline 
   $\mathcal{P}_3$ & $\left(1-3\omega,0,0,0,2-3\omega\right)$ & \multicolumn{1}{|p{4cm}|}{\centering Stable for $n>\frac{5}{4}$,  $\frac{2}{3}<\omega<\frac{4n}{3}-1$, \\ Unstable for $n<\frac{5}{4}$,  $\frac{4n}{3}-1<\omega<\frac{2}{3}$, \\ Non-hyperbolic or saddle otherwise} & $a\sim(-t)^{1/2}$  & $R_*=0$ & \multicolumn{1}{|p{3cm}|}{\centering Decreasing for \\ $\omega\geq\frac{2}{3}$, \\ Increasing for \\ $\omega<\frac{2}{3}$.}\\
   \hline 
   $\mathcal{P}_4$ & \multicolumn{1}{|p{4cm}|}{\centering \begin{small}$\left(- \frac{3(1+\omega)(n-1)}{n},2-\frac{3(1+\omega)}{2n}, \right.$ \\ $\left. \frac{2}{n}-\frac{3(1+\omega)}{2n^2},0, \right.$ \\ $\left. \scriptscriptstyle \left(\frac{2}{n}-1\right) - \frac{3}{2}(1+\omega)\left(\frac{1}{n}-1\right)\left(\frac{1}{n}-2\right)\right)$ \end{small}} & \multicolumn{1}{|p{3cm}|}{\centering Depends on \\ model parameters} & $a\sim(-t)^\frac{2n}{3(\omega+1)}$ & \multicolumn{1}{|p{2cm}|}{\centering $R_*\gtreqless0$ for \\ $\omega\lesseqgtr\frac{4n}{3}-1$} & \multicolumn{1}{|p{3cm}|}{\centering Decreasing for \\ $\omega\geq\frac{1}{2n-1}$, \\ Increasing for \\ $\omega<\frac{1}{2n-1}$.}\\
   \hline 
\end{tabular}}
    \caption{Isotropic equilibrium points for $R^n$ gravity ($n>1$) in presence of an isotropic fluid. The physical viability of the solution corresponding to an equilibrium point can be guaranteed when $R_* = 6H_*^{2}u_{2*}>0$, where the subscript $*$ implies values calculated at the equilibrium point. The behaviour of small anisotropy is investigated via Eq. \eqref{isotropisation}. The conclusion regarding the point $\mathcal{P}_2$ as written in the caption of Tab. \ref{tab:fp_mon_aniso} holds in this case as well.}
    \label{tab:fp_mon_iso}
\end{table}
\begin{table}[h]
\resizebox{\textwidth}{!}{
    \centering
    \begin{tabular}{|c|c|c|c|c|c|}
   \hline
   Point & \multicolumn{1}{p{5cm}}{\centering Coordinates \\ $(u_1,u_2,u_3,u_4)$} & Stability & Cosmology & Sign of $R_*$ & \multicolumn{1}{|p{3cm}|}{\centering Behavior of small\\ anisotropy} \\
   \hline
    $\mathcal{P}_1$ &  $\left(-1,0,0,0\right)$ & Non-hyperbolic & $a\sim(-t)^{1/2}$ & $R_*=0$ & Decreasing \\
    \hline
    $\mathcal{P}_2$ & $\left(\frac{2(n-2)}{2 n-1},\frac{n(4n-5)}{(n-1)(2n-1)},\frac{(4n-5)}{(n-1)(2n-1)},0\right)$ & \multicolumn{1}{|p{2cm}|}{\centering Stable for $1<n<\frac{5}{4}$, \\ Unstable for $n>\frac{5}{4}$, \\ Saddle for $n=\frac{5}{4}$.} & $a\sim(-t)^{\frac{2n^2 - 3n +1}{2-n}}$  & $R_*\gtreqless0$ for $n\gtreqless\frac{5}{4}$ & \multicolumn{1}{|p{3cm}|}{\centering Decreasing for \\ $1<n\leq\frac{5}{4}$, \\ Increasing for \\$n>\frac{5}{4}$.}  \\
    \hline
\end{tabular}}
    \caption{Isotropic equilibrium points for $R^n$ gravity ($n>1$) in absence of any fluid. The physical viability of the solution corresponding to an equilibrium point can be guaranteed when $R_* = 6H_*^{2}u_{2*}>0$, where the subscript $*$ implies values calculated at the equilibrium point. The behaviour of small anisotropy is investigated via Eq. \eqref{isotropisation}. The conclusion regarding the point $\mathcal{P}_2$ as written in the caption of Tab. \ref{tab:fp_mon_aniso} holds in this case as well.}
    \label{tab:fp_mon_vac}
\end{table}
\subsection{Second example: $f(R) = R + \alpha R^{2}$, $(\alpha>0)$}
Our next example is the quadratic gravity $f(R) = R + \alpha R^2$ with $\alpha>0$, a model made famous by Starobinsky's curvature driven inflation model \cite{Starobinsky:1980te}. Absence of tachyonic instability requires $f''(R)>0$, which implies $\alpha>0$. Therefore tachyonic instability is not an issue for this model \footnote{The case $\alpha<0$, although plagued by tachyonic instability, can give rise to nonsingular bounces. See Ref. \cite{Paul:2014cxa}.}. The absence of ghost instability requires $f'>0$, which requires
\begin{equation}
    1 + 2\alpha R > 0.
\end{equation}
For this $f(R)$ model $\Gamma,\,h\left(\frac{u_2}{u_3}\right)$ and $l(u_1,u_2,u_4)$ are as follows:
\begin{equation}
\Gamma = \frac{1+2\alpha R}{2 \alpha R}, \quad h(u_2,u_3) = \frac{u_{2}}{2 \left(u_{2}-u_{3}\right)}, \quad
l(u_1,u_2,u_4) = \frac{u_{2}}{2\left(u_{1}-u_{4}^2 +1 \right)}
\end{equation}
The isotropic equilibrium points in presence of anisotropic fluid, isotropic fluid and in absence of any matter component are listed in tables \ref{tab:fp_quad_aniso}, \ref{tab:fp_quad_iso} and \ref{tab:fp_quad_vac} respectively.
\begin{table}[H]
    \centering
    \resizebox{\textwidth}{!}{
    \begin{tabular}{|c|c|c|c|c|c|}
   \hline
   Point & $(u_1,u_2,u_3,u_4^\pm,u_5)$ & Stability & Cosmology & \multicolumn{1}{|p{3cm}|}{\centering Condition for \\ physical viability} & \multicolumn{1}{|p{3cm}|}{\centering Behavior of small \\ anisotropy} \\
   \hline
    $\mathcal{Q}_1$ & $\left(0,2,1,0,0\right)$ & Non-hyperbolic & $a \sim e^{c(-t)}$ ($c>0$) & Always viable & Increasing \\
    \hline
    $\mathcal{Q}_2$ & $\left(4,0,-5,0,0\right)$ & Saddle & $a \sim \left(-t\right)^{1/2}$ & Always viable & Increasing \\
    \hline
\end{tabular}}
    \caption{Isotropic equilibrium points for $R + \alpha R^2$ gravity ($\alpha>0$) in presence of an anisotropic fluid. Physical viability of the solution corresponding to an equilibrium point requires the absence of ghost instability, i.e. $1 + 2\alpha R_*>0$, where the subscript $*$ implies values calculated at the equilibrium point. The behaviour of small anisotropy is investigated via Eq. \eqref{isotropisation}. None of the equilibrium points represents an isotropising contracting cosmology.}
    \label{tab:fp_quad_aniso}
\end{table}
\begin{table}[H]
\resizebox{\textwidth}{!}{
    \centering
    \begin{tabular}{|c|c|c|c|c|c|}
   \hline
   Point &   $(u_1,u_2,u_3,u_4,u_5)$ & Stability & Cosmology & \multicolumn{1}{|p{3cm}|}{\centering Condition for \\ physical viability} & \multicolumn{1}{|p{3cm}|}{\centering Behavior of small \\ anisotropy} \\
   \hline
     $\mathcal{Q}_1$ & $\left(0,2,1,0,0\right)$ & Non-hyperbolic & $a \sim e^{c(-t)}$ ($c>0$) & Always viable & Increasing \\
    \hline
    $\mathcal{Q}_2$ & $\left(4,0,-5,0,0\right)$ & Saddle & $a \sim \left(-t\right)^{1/2}$ & Always viable & Increasing \\
    \hline
      $\mathcal{Q}_3$ & \multicolumn{1}{|p{4.3cm}|}{\centering $\left(-\frac{3}{2}\left(1+\omega\right),\frac{1}{4}\left(5-3\omega\right),\right.$ \\ $      \left.\frac{1}{8}\left(5-3\omega\right)  ,0,-\frac{9}{8}\left(1+\omega\right)\right)$} & \multicolumn{1}{|p{3cm}|}{\centering Stable for $\omega>\frac{1}{3}+\frac{8\sqrt{6}}{9}$, \\ Non-hyperbolic or saddle otherwise.} & $a \sim \left(-t\right)^\frac{4}{3(1+\omega)}$ & $\omega<-1$ & \multicolumn{1}{|p{3cm}|}{\centering Decreasing for $\omega\geq\frac{1}{3}$, \\ Increasing for $\omega<\frac{1}{3}$.}\\
    \hline
\end{tabular}}
    \caption{Isotropic equilibrium points for $R + \alpha R^2$ gravity ($\alpha>0$) in presence of an isotropic fluid. Physical viability of the solution corresponding to an equilibrium point requires the absence of ghost instability, i.e. $1 + 2\alpha R_*>0$ as well as $u_{5*}>0$, where the subscript $*$ implies values calculated at the equilibrium point. The behaviour of small anisotropy is investigated via Eq. \eqref{isotropisation}. The isotropic equilibrium point $\mathcal{Q}_3$ can represent a stable contracting power-law cosmology with isotropisation when $\omega>\frac{1}{3}+\frac{8\sqrt{6}}{9}$, but in that case, the solution does not remain physically viable anymore.}
    \label{tab:fp_quad_iso}
\end{table}
\begin{table}[H]
    \centering
     \resizebox{\textwidth}{!}{
       \begin{tabular}{|c|c|c|c|c|c|}
   \hline
   Point & $(u_1,u_2,u_3,u_4)$ & Stability & Cosmology & \multicolumn{1}{|p{3cm}|}{\centering Condition for \\ physical viability} & \multicolumn{1}{|p{3cm}|}{\centering Behavior of small \\ anisotropy} \\
   \hline
    $\mathcal{Q}_1$ & $\left(0,2,1,0\right)$ & Non-hyperbolic & $a \sim e^{c(-t)}$ ($c>0$) & Always viable & Increasing \\
    \hline 
     $\mathcal{Q}_2$ & $\left(4,0,-5,0,0\right)$ & Saddle & $a \sim \left(-t\right)^{1/2}$ & Always viable & Increasing \\
    \hline
\end{tabular}}
    \caption{Isotropic equilibrium points for $R + \alpha R^2$ gravity ($\alpha>0$) in absence of any fluid. Physical viability of the solution corresponding to an equilibrium point requires the absence of ghost instability, i.e. $1 + 2\alpha R_*$, where the subscript $*$ implies values calculated at the equilibrium point. The behaviour of small anisotropy is investigated via Eq. \eqref{isotropisation}. None of the equilibrium points represents an isotropising contracting cosmology.}
    \label{tab:fp_quad_vac}
\end{table}
\subsection{Third example: $f(R) = \frac{1}{\alpha}e^{\alpha R}$, $(\alpha>0)$}
Lastly, we consider the exponential form $f(R)=\frac{1}{\alpha}e^{\alpha R}$ with $\alpha>0$. This $f(R)$ has been considered previously in the context of both nonsingular bounce and inflation \cite{Abdelwahab:2007jp,Bari:2018aac}. The advantage of this particular form is that $f'$ and $f''$ are always positive. So even if $R$ becomes highly negative, which might be the case for, say, an isotropic ekpyrotic contraction with $0<\alpha<\frac{1}{2}$, one need not worry about a ghost or tachyonic instability. For this $f(R)$ model $\Gamma,\,h\left(\frac{u_2}{u_3}\right)$ and $l(u_1,u_2,u_4)$ are as follows:
\begin{equation}
\Gamma = \frac{1}{\alpha R}, \quad h(u_2,u_3) = \frac{u_{3}}{u_{2}}, \quad
l(u_1,u_2,u_4) = \frac{u_{2}+u_{4}^2-u_{1}-1}{u_{2}}.
\end{equation}
The isotropic equilibrium points in presence of anisotropic fluid, isotropic fluid and in absence of any matter component, along with their nature of stability and the conditions for the absence of ghost and tachyonic instability, are listed in tables \ref{tab:fp_exp_aniso}, \ref{tab:fp_exp_iso} and \ref{tab:fp_exp_vac} respectively.
\begin{table}[H]
    \centering
    \resizebox{\textwidth}{!}{
    \begin{tabular}{|c|c|c|c|c|}
   \hline
   Point & Coordinates $(u_1,u_2,u_3,u_4^\pm,u_5)$ & Stability & Cosmology & Behaviour of small anisotropy \\
   \hline
    $\mathcal{S}_1$ & $\left( -1,0,0,0,0\right)$& Non-hyperbolic & $a \sim \left(-t\right)^{1/2}$ & Decreasing \\
    \hline
    $\mathcal{S}_2$ & $\left(1,2,0,0,0\right)$& Non-hyperbolic & $a \sim e^{c(-t)}$ ($c>0$) & Increasing \\
    \hline
    $\mathcal{S}_3$ & $\left(0,2,1,0,0\right)$& Saddle & $a \sim e^{c(-t)}$ ($c>0$) & Increasing \\
    \hline
\end{tabular}}
    \caption{Isotropic equilibrium points for $\frac{1}{\alpha}e^{\alpha R}$ gravity ($\alpha>0$) in presence of an anisotropic fluid. The behaviour of small anisotropy is investigated via Eq. \eqref{isotropisation}. }
    \label{tab:fp_exp_aniso}
\end{table}
\begin{table}[H]
    \centering
 \resizebox{\textwidth}{!}{   
 \begin{tabular}{|c|c|c|c|c|}
   \hline
   Point & Coordinates $(u_1,u_2,u_3,u_4,u_5)$ & Stability & Cosmology & Behaviour of small anisotropy \\
   \hline
   $\mathcal{S}_1$ & $\left( -1,0,0,0,0\right)$ & Non-hyperbolic & $a \sim \left(-t\right)^{1/2}$ & Decreasing \\
    \hline
   $\mathcal{S}_2$ & $\left(1,2,0,0,0\right)$& Non-hyperbolic & $a \sim e^{c(-t)}$ ($c>0$) & Increasing \\
    \hline
    $\mathcal{S}_3$ & $\left(0,2,1,0,0\right)$& Saddle & $a \sim e^{c(-t)}$ ($c>0$) & Increasing \\
    \hline
     $\mathcal{S}_4$ & $\left(1-3\omega,0,0,0,2-3\omega\right)$& \multicolumn{1}{|p{2cm}|}{\centering  Stable for $\omega>\frac{2}{3}$, \\ Saddle for $\omega<\frac{2}{3}$, \\ Non-hyperbolic otherwise.} & $a \sim \left(-t\right)^{1/2}$ & \multicolumn{1}{|p{4cm}|}{\centering Decreasing for \\ $\omega\geq\frac{2}{3}$, \\ Increasing for \\ $\omega<\frac{2}{3}$.} \\
    \hline
     $\mathcal{S}_5$ & $\left(-3(1+\omega),2,0,0,-4-3\omega \right)$& Non-hyperbolic & $a \sim e^{c(-t)}$ ($c>0$) & \multicolumn{1}{|p{4cm}|}{\centering Decreasing for \\ $\omega\geq0$, \\ Increasing for \\ $\omega<0$.} \\
    \hline
\end{tabular}}
    \caption{Isotropic equilibrium points for $\frac{1}{\alpha}e^{\alpha R}$ gravity ($\alpha>0$) in presence of an isotropic fluid. The behaviour of small anisotropy is investigated via Eq. \eqref{isotropisation}. The isotropic equilibrium point $\mathcal{S}_4$ can represent a stable contracting power-law cosmology with isotropisation for $\omega>\frac{2}{3}$, but in that case, $u_5$ becomes negative at this point, rendering the solution not physically viable.}
    \label{tab:fp_exp_iso}
\end{table}
\begin{table}[H]
    \centering
    \resizebox{\textwidth}{!}{
    \begin{tabular}{|c|c|c|c|c|}
   \hline
   Point & Coordinates $(u_1,u_2,u_3,u_4)$ & Stability & Cosmology & Behaviour of small anisotropy \\
   \hline
     $\mathcal{S}_1$ & $\left( -1,0,0,0\right)$ & Non-hyperbolic & $a \sim \left(-t\right)^{1/2}$ & Decreasing \\
    \hline 
    $\mathcal{S}_2$ & $\left( 1,2,0,0\right)$ & Non-hyperbolic & $a \sim e^{c(-t)}$ ($c>0$) & Increasing \\
    \hline
     $\mathcal{S}_3$ & $\left(0,2,1,0\right)$ & Saddle & $a \sim e^{c(-t)}$ ($c>0$) & Increasing \\
    \hline 
\end{tabular}}
    \caption{Isotropic equilibrium points for $\frac{1}{\alpha}e^{\alpha R}$ gravity ($\alpha>0$) in absence of any fluid. The behaviour of small anisotropy is investigated via Eq. \eqref{isotropisation}.}
    \label{tab:fp_exp_vac}
\end{table}

\section{A form-independent analysis}\label{sec:form-independent}

In this section, we do not specify the $f(R)$ function a priori, but we work backwards, starting from an ansatz for the cosmological evolution. Therefore, we present a form-independent analysis. The motivation behind this analysis is that even if one gets a contracting ekpyrotic solution as an equilibrium point in any given $f(R)$ model, there is no guarantee that it will be connected to a subsequent nonsingular bouncing phase via a heteroclinic phase trajectory. Therefore, we try to alleviate this issue in this section.

Let us consider an ansatz for a cosmological evolution as given by the Hubble parameter evolution: 
\begin{align}\label{ansatz}
H(t)= \frac{\beta t}{\alpha^2 + t^2},
\end{align}
where $-\infty < t < \infty$, $\alpha$ being a characteristic time-scale, i.e., $[\alpha]=[t]$, and $\beta$ is a dimensionless (stretching) constant. This ansatz has the following properties
\begin{itemize}
    \item When $|t|\gg\alpha$, $H(t)\approx\frac{\beta}{t}$ i.e. $a(t)\sim|t|^\beta$.\\
    \item When $|t|\ll\alpha$, $H(t)\approx \beta t/\alpha^2$ i.e. $a(t)\sim e^{\frac{\beta}{2 \alpha^2}t^2}$.\\
\end{itemize}
For simplicity, $\beta$ is taken of the same value of $\alpha$,  and assume  $\alpha, \beta\in (0,1)$, where $\alpha$ is the relevant parameter.

This ansatz specifies an ekpyrotic contraction phase followed by a nonsingular bounce, with a smooth transition. One can, of course, try to reconstruct the $f(R)$ that can give rise to such an evolution, but the reconstruction method does not give a compact, functional form that may help in our subsequent investigation of its isotropisation property. We, therefore, do not follow that route. Rather, the question we address is, given that such an $f(R)$ exists \footnote{In fact, given any $a(t)$, one can always, in principle, reconstruct the $f(R)$, even though rarely does this gives any compact, functional form.}
\begin{itemize}
    \item Does it give rise to isotropic equilibrium points?
    \item What qualitative conclusion can we draw about the evolution of small metric anisotropy as the universe goes from an ekpyrotic contraction phase to a nonsingular bounce?
    \item Does the $f(R)$ remain physically viable throughout the evolutionary history we are interested in?
\end{itemize}
To this goal, we utilise an alternative formulation of a dynamical system in $f(R)$ gravity presented in Ref. \cite{Chakraborty:2021jku}. The formulation made use of the deceleration and jerk parameter
\begin{subequations}
\begin{eqnarray}
&& q = - \frac{1}{aH^2}\frac{d^{2}a}{dt^2} = - 1 - \frac{\dot{H}}{H^2},\\
&& j = \frac{1}{aH^3}\frac{d^{3}a}{dt^3} = - 2 - 3q + \frac{\ddot{H}}{H^3}.
\end{eqnarray}
\end{subequations}
There is an important relation given between the two parameters \footnote{The reader might have encountered the relation $j = q(2q+1)-\frac{dq}{dN}$ ($N=\ln a$) that is very common in literature involving cosmography. The $`-'$ sign in the last term comes because $\tau=-N$.} 
\begin{align}\label{cosm}
    j = q(2q+1) + \frac{dq}{d\tau}.
\end{align}
For the ansatz \eqref{ansatz}, the deceleration is
\begin{equation}\label{ansatz_q}
 q(t) = - 1 - \frac{\alpha^2 - t^2}{\alpha t^2},   
\end{equation}
which can be inverted to write 
\begin{eqnarray}
 t^2 = \frac{\alpha^2}{1 - \alpha(q+1)}.
\end{eqnarray}
That gives
\begin{equation}\label{ansatz_cosm}
j(t(q)) = \frac{\alpha-2}{\alpha^2}\left[2-\alpha \left(2+3q\right) \right] .   
\end{equation}
That is the cosmographic constraint corresponding to time evolution in Eq. \eqref{ansatz}. Therefore, it is essential to mention that although the cosmological solution in Eq. \eqref{ansatz} necessarily corresponds to the cosmographic constraint \eqref{ansatz_cosm}, the inverse is not valid. That is because, when written in terms of $H$ and its derivatives, the constraint \eqref{ansatz_cosm} is a second-order differential equation in $H(t)$. The general solution to this equation is a much bigger family, of which the particular solution \eqref{ansatz} is just one part.

The next step is to write the dynamical equations. First, one can notice that the term $\Gamma$ appears in the dynamical system only in the combination $u_1 u_2 \Gamma$, which can be rewritten in terms of cosmographic parameters as
\begin{equation}\label{u1_u2_Gamma}
   u_1 u_2 \Gamma = - 2 u_{4}^{2}(3+u_1) + u_{5}\left(3u_4^{+}\mu_{+} + \sqrt{3} u_4^{-} \mu_{-}\right) + j - q - 2.  
\end{equation}
One can write an extended dynamical system consisting of the following dynamical equations
\begin{subequations}
\begin{align}
\frac{d u_1}{d\tau}&  = u_1 \left(u_1+ u_2-u_4^2 \right)-u_2 + 3 u_3+u_{4}^2 + 3 \omega u_5 -1,\\
\frac{d u_2}{d\tau}& = 2u_2 \left(u_2- u_4^2-2\right)+ 2 u_{4}^{2}(3+u_1) - u_{5}\left(3 u_4^{+} \mu_{+} + \sqrt{3} u_4^{-} \mu_{-}\right) \nonumber \\
& - \left(\frac{2-\alpha}{\alpha}\right) \left(3 q - \frac{2 (1-\alpha)}{\alpha} \right) +q + 2,\\
 \frac{d u_3}{d\tau}& = u_3 \left(u_1+2 u_2-2 u_4^2-4\right)+2 u_{4}^{2}(3+u_1) - u_{5}\left(3 u_4^{+} \mu_{+} + \sqrt{3} u_4^{-} \mu_{-}\right) \nonumber \\
& - \left(\frac{2-\alpha}{\alpha}\right) \left(3 q + \frac{2 (1-\alpha)}{\alpha} \right) + q + 2, \\
\frac{d u_{4}^+}{d\tau}& =\frac{3}{2} \mu _+
   \left(-u_1+u_2-u_3+u_4^2-1\right)+u_4^{+}
   \left(u_1+u_2-u_4^2+1\right),\\
\frac{d u_{4}^-}{d\tau}& =\frac{\sqrt{3} }{2} \mu _-
   \left(-u_1+u_2-u_3+u_4^2-1\right)+ u_4^{-}
   \left(u_1+u_2-u_4^2+1\right),\\
\frac{d u_5}{d\tau}& = u_5 \left(3 \omega -1 + u_1 + 2u_2 - 2 u_{4}^2 + 3 \mu_{+} u_{4}^{+}+ \sqrt{3} \mu_{-} u_{4}^-\right), \\
\frac{dq}{d\tau}&  = - 2\left(q-\frac{1-\alpha}{\alpha}\right) \left(q-\frac{2-\alpha}{\alpha}\right),    
\end{align}
\end{subequations}
along with two constraint equations. They are provided by the Friedmann constraint and the definition of the Ricci scalar, respectively
\begin{eqnarray}\label{const_fried_ricci}
1 + u_1 - u_2 + u_3 - u_4^2 - u_5 = 0, \quad u_2 = 1 - q + u_4^2.
\end{eqnarray}
These two constraints can be used to eliminate two of the variables. We choose to eliminate $u_2$ and $u_3$. However, it is important to mention that not all the available phase space is physically viable. The conditions for the absence of ghost and tachyonic instability restrict the physically viable region of the phase space. One can write the physical viability condition to be
\begin{equation}\label{phys_viab}
  \frac{1}{u_2 \Gamma} = \frac{6f''H^2}{f'} = 
   \frac{u_1}{- 2 u_{4}^{2}(3+u_1) + u_{5}\left(3 u_4^{+} \mu_{+} + \sqrt{3} u_4^{-} \mu_{-}\right) + j -q - 2} > 0.
\end{equation}
The reduced dynamical system for the cases of anisotropic fluid, isotropic fluid and vacuum are written below. 
\begin{itemize}
    \item Anisotropic fluid:
    \begin{subequations}
    \begin{eqnarray}
    && \frac{d u_1}{d\tau} = u_{1}\left(u_{1}-2-q\right)-2 q +6 u_{4}^2 + 3 (1+\omega)u_{5}-2,\\
    && \frac{d u_{4}^+}{d\tau} = -\frac{3}{2} \mu_{+} u_{5} + u_{4}^{+}\left(2-q+u_{1}\right),\\
    && \frac{d u_{4}^-}{d\tau} =-\frac{\sqrt{3}}{2} \mu_{-} u_{5} + u_{4}^{-}\left(2-q+u_{1}\right),\\
    && \frac{d u_{5}}{d\tau} = u_{5} \left(1-2q+u_{1}+3\omega +3 \mu_{+} u_{4}^{+}+\sqrt{3}\mu_{-}u_{4}^{-}\right),\\
    && \frac{d q}{d\tau} = - 2\left(q-\frac{1-\alpha}{\alpha}\right) \left(q-\frac{2-\alpha}{\alpha}\right). 
    \end{eqnarray}
    \end{subequations}
    \item Isotropic fluid:
    \begin{subequations}
    \begin{eqnarray}
    && \frac{d u_1}{d\tau} = u_{1}\left(u_{1}-2-q\right)-2 q +6 u_{4}^2 + 3 (1+\omega)u_{5}-2,\\
    && \frac{d u_{4}}{d\tau} = u_{4} \left(2-q+u_{1}\right),\\
    && \frac{d u_{5}}{d\tau} = u_{5} \left(1-2q+u_{1}+3\omega \right),\\
    && \frac{d q}{d\tau} = - 2\left(q-\frac{1-\alpha}{\alpha}\right) \left(q-\frac{2-\alpha}{\alpha}\right).
    \end{eqnarray}
    \end{subequations}
    \item Vacuum: 
    \begin{subequations}
    \begin{eqnarray}
    && \frac{du_1}{d\tau} = u_{1}\left(u_{1}-2-q \right) -2q+6 u_{4}^2 -2,\\
    && \frac{du_{4}}{d\tau}  = u_{4} \left(2-q + u_{1} \right),\\
    && \frac{d q}{d\tau} = - 2\left(q-\frac{1-\alpha}{\alpha}\right) \left(q-\frac{2-\alpha}{\alpha}\right).
    \end{eqnarray}
    \end{subequations}
    \end{itemize}
In each case, we also list the isotropic equilibrium points along with their nature of stability as obtained by a Jacobian analysis in tables \ref{tab:fp_aniso}, \ref{tab:fp_iso}, \ref{tab:fp_vac} respectively. Jacobian eigenvalues are given in appendix \ref{app:compactification} for the reader's reference. Due to the second constraint equation in Eq. \eqref{const_fried_ricci}, that exact condition \eqref{exact-isotropisation} for the stability of the ``isotropic invariant submanifold'' becomes
\begin{equation}\label{isotropisation_new}
   \left(u_1 -q + 2\right)<0.
\end{equation}
Since we have chosen to trade off $u_2$ in terms of $q$ in our analysis, the condition that small perturbative anisotropy will die out for an isotropic equilibrium point is now expressed as \eqref{isotropisation_new} instead of Eq. \eqref{isotropisation}. This condition, again, is found from the $u_4$-equation(s) demanding that the submanifold $u_4=0$ is attracting for initial $u_4>0$. However, when $\left(u_1 -q + 2\right)=0$, Eq. \eqref{isotropisation_new} is inconclusive to determine the behavior of small perturbative anisotropy. Indeed what that means is that small anisotropies remain constant (neither increase nor decrease). For each equilibrium point, we mention the criteria for physical viability obtained from the condition \eqref{phys_viab} and the behaviour of small perturbative anisotropy obtained from the isotropisation condition \eqref{isotropisation_new}.

\begin{table}[H]
\resizebox{\textwidth}{!}{
\centering
\begin{tabular}{|c|c|c|c|c|c|}
\hline
Point & Coordinates $(u_1,u_4^\pm,u_5,q)$ & Stability & Cosmology & Physical viability & \multicolumn{1}{|p{2cm}|}{\centering Behaviour of small \\ anisotropy} \\
\hline
$\mathcal{K}_{1-}$ & $\left(\frac{1+\alpha- \sqrt{\alpha ^2+10 \alpha +1}}{2 \alpha },0,0, \frac{1-\alpha}{\alpha}\right)$  & Saddle & $a\sim \left(-t\right)^{\alpha}$ &  $\frac{1}{2}<\alpha <1$ & \multicolumn{1}{|p{3cm}|}{\centering Decreasing for \\ $0<\alpha <\frac{1}{2}$, \\ Increasing for\\ $\frac{1}{2}<\alpha <1$, \\ Inconclusive for \\ $\alpha=\frac{1}{2}$.} \\
\hline
$\mathcal{K}_{1+}$ & $\left(\frac{1+\alpha+ \sqrt{\alpha ^2+10 \alpha +1}}{2 \alpha },0,0, \frac{1-\alpha}{\alpha}\right)$ &  \multicolumn{1}{|p{4.5cm}|}{\centering Unstable for $0<\alpha<\frac{1}{2}$,  $\omega>\frac{3-7\alpha}{6\alpha}-\frac{1}{6} \sqrt{\frac{\alpha^2 +10\alpha +1}{\alpha^2}}$, \\ Saddle otherwise.} & $a\sim \left(-t\right)^{\alpha}$ & $0<\alpha <\frac{1}{2}$ & \multicolumn{1}{|p{2cm}|}{\centering Increasing for \\ $\alpha>0$}\\
\hline
$\mathcal{K}_{2-}$ &  $\left(\frac{2+\alpha- \sqrt{\alpha ^2+20 \alpha +4}}{2 \alpha },0,0, \frac{2-\alpha}{\alpha}\right)$ & \multicolumn{1}{|p{4.5cm}|}{\centering Stable for $0<\alpha<1$,  $\omega<\frac{6-7\alpha}{6\alpha}+\frac{1}{6} \sqrt{\frac{\alpha^2 +20\alpha +4 }{\alpha^2}}$, \\ Saddle otherwise.} & $a\sim \left(-t\right)^{\alpha/2}$ & \multicolumn{1}{|p{2cm}|}{\centering Not viable \\
for $0<\alpha<1$} & \multicolumn{1}{|p{2cm}|}{\centering Decreasing for \\ $0<\alpha<1$ }\\
\hline
$\mathcal{K}_{2+}$ & $\left(\frac{2+\alpha+ \sqrt{\alpha ^2+20 \alpha +4}}{2 \alpha },0,0, \frac{2-\alpha}{\alpha}\right)$  & Saddle & $a\sim \left(-t\right)^{\alpha/2}$ & $0<\alpha<1$& \multicolumn{1}{|p{2cm}|}{\centering Increasing for \\ $0<\alpha<1$} \\
\hline
\end{tabular}}
\caption{Isotropic equilibrium points in the presence of an anisotropic fluid for $f(R)$ theories that can be reconstructed based on the cosmographic condition \eqref{ansatz_cosm}. Physical viability and behaviour of small anisotropy are investigated via Eq. \eqref{phys_viab} and Eq. \eqref{isotropisation_new} respectively.}
\label{tab:fp_aniso}
\end{table}

\begin{table}[]
\resizebox{\textwidth}{!}{
\centering
\begin{tabular}{|c|c|c|c|c|c|}
\hline
Point & Coordinates $(u_1,u_4,u_5,q)$ & Stability & Cosmology &  \multicolumn{1}{|p{2cm}|}{\centering Physical \\ viability} & \multicolumn{1}{|p{2cm}|}{\centering Behaviour of small \\ anisotropy} \\
\hline
 $\mathcal{K}_{1-}$ & $\left(\frac{1+\alpha- \sqrt{\alpha ^2+10 \alpha +1}}{2 \alpha },0,0, \frac{1-\alpha}{\alpha}\right)$  & Saddle & $a\sim \left(-t\right)^{\alpha}$ &  $\frac{1}{2}<\alpha <1$ & \multicolumn{1}{|p{2cm}|}{\centering Decreasing for $0<\alpha <\frac{1}{2}$, \\ Increasing for $\frac{1}{2}<\alpha<1$, \\ Inconclusive for $\alpha=\frac{1}{2}$.} \\
\hline 
$\mathcal{K}_{1+}$ & $\left(\frac{1+\alpha+ \sqrt{\alpha ^2+10 \alpha +1}}{2 \alpha },0,0, \frac{1-\alpha}{\alpha}\right)$ &  \multicolumn{1}{|p{4.5cm}|}{\centering Unstable for $0<\alpha<\frac{1}{2}$,  $\omega>\frac{3-7\alpha}{6\alpha}-\frac{1}{6} \sqrt{\frac{\alpha^2 +10\alpha +1}{\alpha^2}}$, \\ Saddle otherwise.} & $a\sim \left(-t\right)^{\alpha}$ & $0<\alpha <\frac{1}{2}$ & \multicolumn{1}{|p{2cm}|}{Increasing for $0<\alpha<1$.}\\
\hline
$\mathcal{K}_{2-}$ &  $\left(\frac{2+\alpha- \sqrt{\alpha ^2+20 \alpha +4}}{2 \alpha },0,0, \frac{2-\alpha}{\alpha}\right)$ & \multicolumn{1}{|p{4.5cm}|}{\centering Stable for $0<\alpha<1$,  $\omega<\frac{6-7\alpha}{6\alpha}+\frac{1}{6} \sqrt{\frac{\alpha^2 +20\alpha +4 }{\alpha^2}}$, \\ Saddle otherwise.} & $a\sim \left(-t\right)^{\alpha/2}$ & \multicolumn{1}{|p{2cm}|}{\centering Not viable \\
for $0<\alpha<1$} & \multicolumn{1}{|p{2cm}|}{Decreasing for $0<\alpha<1$.} \\
\hline
$\mathcal{K}_{2+}$ & $\left(\frac{2+\alpha+ \sqrt{\alpha ^2+20 \alpha +4}}{2 \alpha },0,0, \frac{2-\alpha}{\alpha}\right)$  & Saddle & $a\sim \left(-t\right)^{\alpha/2}$ & $0<\alpha<1$ & \multicolumn{1}{|p{2cm}|}{Increasing for $0<\alpha<1$.} \\
\hline
$\mathcal{K}_{3}$ & \multicolumn{1}{|p{4.5cm}|}{\centering $\left(\frac{2}{\alpha}-3 \left(1+\omega\right), 0, \right. $\\ $\left.
\scriptscriptstyle\frac{\alpha  (-3 \alpha  (\omega +1) (3 \omega +4)+9 \omega +13)-2}{3 \alpha ^2 (\omega +1)}, \frac{1-\alpha }{\alpha }\right)$} &  Depends on model parameters & $a\sim \left(-t\right)^{\alpha}$ & \multicolumn{1}{|p{2cm}|}{\centering $ 0<\alpha<\frac{1}{2},\,$ \\ $\,\,\omega<\frac{2-3\alpha}{3\alpha}$ \\ or \\ $ \frac{1}{2}<\alpha<1,\,$ \\ $\,\,\omega>\frac{2-3 \alpha}{3\alpha}$} & \multicolumn{1}{|p{2cm}|}{\centering Decreasing for $\omega<\frac{1-2\alpha}{\alpha}$, \\ Increasing for $\omega>\frac{1-2\alpha}{\alpha}$, \\ Inconclusive for $\omega=\frac{1-2\alpha}{\alpha}$}\\
\hline
$\mathcal{K}_{4}$ & \multicolumn{1}{|p{4.5cm}|}{\centering $\left(\frac{4}{\alpha}-3 \left(1+\omega\right), 0, \right. $\\ $\left.
\scriptscriptstyle \frac{8 (\alpha -1)}{3 \alpha ^2 (\omega +1)}+\frac{6}{\alpha }-3 \omega -4, \frac{2-\alpha }{\alpha }\right)$} & Depends on model parameters & $a\sim \left(-t\right)^{\alpha/2}$ & \multicolumn{1}{|p{2cm}|}{\centering $0<\alpha<1$,  $\omega < \frac{4-3\alpha}{3\alpha}$} & \multicolumn{1}{|p{2cm}|}{\centering Decreasing for $\omega<\frac{2-2\alpha}{\alpha}$, \\ Increasing for $\omega>\frac{2-2\alpha}{\alpha}$, \\ Inconclusive for $\omega=\frac{2-2\alpha}{\alpha}$}\\
\hline
\end{tabular}}
\caption{Isotropic equilibrium points in the presence of an isotropic fluid for $f(R)$ theories that can be reconstructed based on the cosmographic condition \eqref{ansatz_cosm}. Physical viability and behaviour of small anisotropy are investigated via Eq. \eqref{phys_viab} and Eq. \eqref{isotropisation_new} respectively.}
\label{tab:fp_iso}
\end{table}

\begin{table}[h]
\centering
\resizebox{\textwidth}{!}{
\begin{tabular}{|c|c|c|c|c|c|}
\hline
Point & Coordinates $(u_1,u_4,q)$ & Stability & Cosmology & Physical viability & \multicolumn{1}{|p{2cm}|}{\centering  Behaviour of small \\ anisotropy} \\
\hline
$\mathcal{K}_{1-}$ & $\left(\frac{1+\alpha- \sqrt{\alpha ^2+10 \alpha +1}}{2 \alpha },0, \frac{1-\alpha}{\alpha}\right)$  & Saddle & $a\sim \left(-t\right)^{\alpha}$ &  $\frac{1}{2}<\alpha <1$ & \multicolumn{1}{|p{3cm}|}{\centering Decreasing for \\ $0<\alpha <\frac{1}{2}$, \\ Increasing for\\ $\frac{1}{2}<\alpha <1$, \\ Inconclusive for \\ $\alpha=\frac{1}{2}$.} \\
\hline 
$\mathcal{K}_{1+}$ & $\left(\frac{1+\alpha+ \sqrt{\alpha ^2+10 \alpha +1}}{2 \alpha },0, \frac{1-\alpha}{\alpha}\right)$ & Unstable & $a\sim \left(-t\right)^{\alpha}$ & $0<\alpha <\frac{1}{2}$ & \multicolumn{1}{|p{2cm}|}{\centering Increasing for \\ $\alpha>0$}\\
\hline
$\mathcal{K}_{2-}$ & $\left(\frac{2+\alpha- \sqrt{\alpha ^2+20 \alpha +4}}{2 \alpha },0, \frac{2-\alpha}{\alpha}\right)$ & Stable & $a\sim \left(-t\right)^{\alpha/2}$ & \multicolumn{1}{|p{2cm}|}{\centering Not viable \\
for $0<\alpha<1$} & \multicolumn{1}{|p{2cm}|}{\centering Decreasing for \\ $0<\alpha<1$ }\\
\hline
$\mathcal{K}_{2+}$ & $\left(\frac{2+\alpha+ \sqrt{\alpha ^2+20 \alpha +4}}{2 \alpha },0, \frac{2-\alpha}{\alpha}\right)$  & Saddle & $a\sim \left(-t\right)^{\alpha/2}$ & $0<\alpha<1$& \multicolumn{1}{|p{2cm}|}{\centering Increasing for \\ $0<\alpha<1$} \\
\hline
\end{tabular}}
\caption{Isotropic equilibrium points in the absence of any fluid for $f(R)$ theories that can be reconstructed based on the cosmographic condition \eqref{ansatz_cosm}. Physical viability and behaviour of small anisotropy are investigated via Eq. \eqref{phys_viab} and Eq. \eqref{isotropisation_new} respectively. Physical viability and behaviour of small anisotropy are investigated via Eq. \eqref{phys_viab_comp} and Eq. \eqref{isotropisation_comp} respectively.}
\label{tab:fp_vac}
\end{table}

One can note the existence of two invariant submanifolds given by
\begin{equation}\label{q-inv}
 q = \frac{1-\alpha}{\alpha}, \quad \frac{2-\alpha}{\alpha}.   
\end{equation}
 These two invariant submanifolds divide the entire phase space into three disjoint regions
\begin{equation}
    -\infty<q<\frac{1-\alpha}{\alpha},\,\frac{1-\alpha}{\alpha}<q<\frac{2-\alpha}{\alpha},\,\frac{2-\alpha}{\alpha}<q<\infty.
\end{equation}
For the particular ansatz we have considered in Eq. \eqref{ansatz}, the bounce occurs at $t=0$. As the universe approaches from the ekpyrotic contraction phase at $t\ll-\alpha$ to the bounce at $t\to0-$, $q$ goes from $\frac{1-\alpha}{\alpha}$ to $-\infty$, as is clear from the expression of the deceleration parameter $q(t)$ in Eq. \eqref{ansatz_q}. Therefore it is the region $-\infty<q<\frac{1-\alpha}{\alpha}$ of the phase space in which the phase trajectories correspond to a smooth transition from an ekpyrotic contraction to a nonsingular bounce. The other two disjoint regions still correspond to contracting cosmology obeying the same cosmographic condition \eqref{ansatz_cosm}, but they do not represent an ekpyrotic phase followed by a nonsingular bounce.
\subsection{The isotropic vacuum submanifold}
One can note that the isotropic vacuum submanifold, given by $(u_4,u_5) = (0,0)$ is always an invariant submanifold. This invariant submanifold is a plane on which the following 2-dimensional dynamical system gives the phase flow
\begin{subequations}\label{dynsys_iso_vac}
    \begin{eqnarray}
    && \frac{du_1}{d\tau} = u_{1}\left(u_{1}-2-q \right) - 2q - 2,\\
    && \frac{dq}{d\tau} = -2\left(q-\frac{1-\alpha}{\alpha}\right) \left(q-\frac{2-\alpha}{\alpha}\right). 
    \end{eqnarray}
\end{subequations}
\begin{figure}[H]
    \centering
    \subfigure[\label{fig:iso_vac_0.2}]{\includegraphics[width=0.45\linewidth]{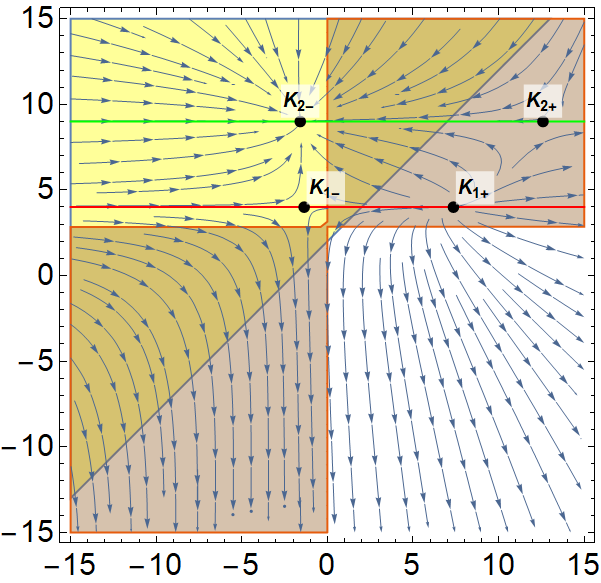}}
    \hspace{0.5cm}
    \subfigure[\label{fig:iso_vac_0.4}]{\includegraphics[width=0.45\linewidth]{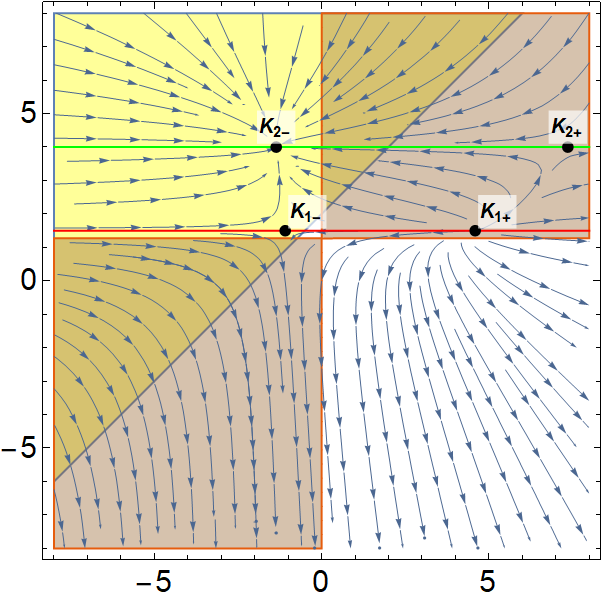}}
    \hspace{0.5cm}
    \subfigure[\label{fig:iso_vac_0.6}]{\includegraphics[width=0.45\linewidth]{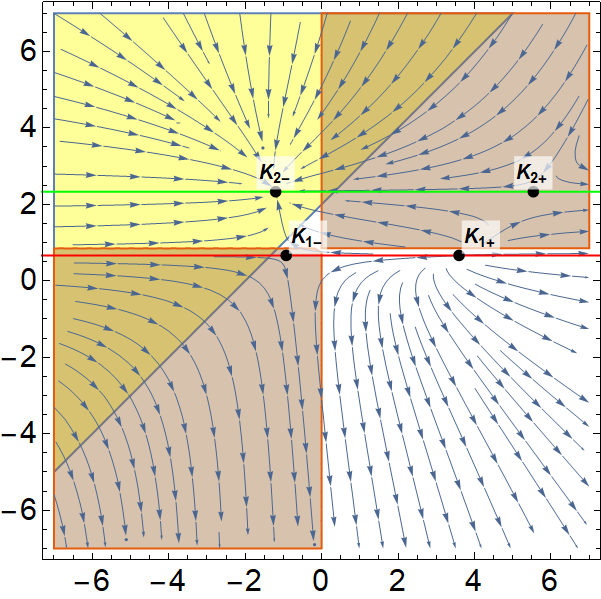}}
    \hspace{0.5cm}
    \subfigure[\label{fig:iso_vac_0.8}]{\includegraphics[width=0.45\linewidth]{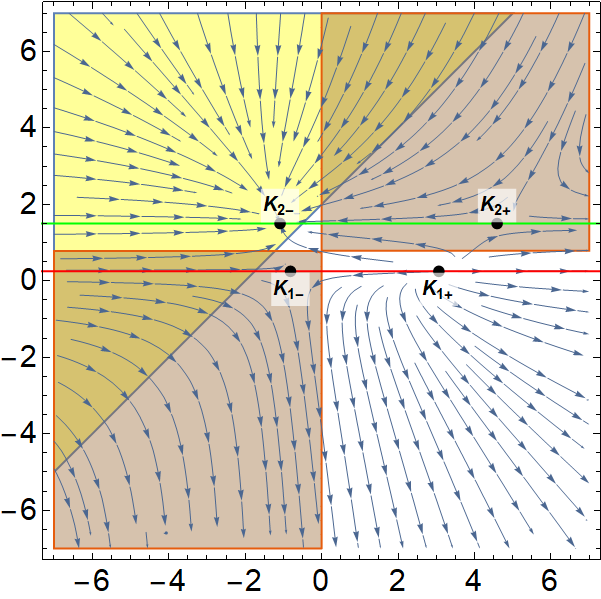}}
    \caption{\label{fig:iso_vac} The phase portrait on the isotropic vacuum invariant submanifold as given by the 2-dimensional dynamical system in Eq. \eqref{dynsys_iso_vac} for \ref{fig:iso_vac_0.2} $\alpha=0.2$, \ref{fig:iso_vac_0.4} $\alpha=0.4$, \ref{fig:iso_vac_0.6} $\alpha=0.6$, \ref{fig:iso_vac_0.8} $\alpha=0.8$. The purple shaded region corresponds to the region of physical viability as given by the condition \eqref{phys_viab}. The yellow shaded region corresponds to the region of isotropisation as given by the condition \eqref{isotropisation_new}.}
  \end{figure}
Figs. \ref{fig:iso_vac} present the phase portrait on the isotropic vacuum invariant submanifold as given by the 2-dimensional dynamical system in Eq. \eqref{dynsys_iso_vac} for \ref{fig:iso_vac_0.2} $\alpha=0.2$, \ref{fig:iso_vac_0.4} $\alpha=0.4$, \ref{fig:iso_vac_0.6} $\alpha=0.6$, \ref{fig:iso_vac_0.8} $\alpha=0.8$. The purple shaded region corresponds to the region of physical viability as given by the condition \eqref{phys_viab}. The yellow shaded region corresponds to the region of isotropisation as given by the condition \eqref{isotropisation_new}.

All the phase trajectories in the figures represent an isotropic vacuum contracting cosmology obeying the cosmographic condition \eqref{ansatz_cosm}. The two invariant submanifolds given by Eq. \eqref{q-inv} are now horizontal lines, and the cosmology representing an ekpyrotic contraction phase smoothly transiting to a nonsingular bouncing phase is represented by the bottom third of the phase portraits, i.e. the portion below the line $q=\frac{1-\alpha}{\alpha}$.

Phase portrait analysis helps analyse the behaviour of small anisotropy and whether the cosmology remains physically viable for this evolutionary history, starting from the ekpyrotic contraction phase up to the nonsingular bounce so that we present a compact phase space. The $u$-coordinate can be compactified via the usual Poincare compactification by introducing the new compact variable $\bar{u}_1$ as
\begin{equation}\label{def_ubar}
    \bar{u}_1 = \frac{u_1}{\sqrt{1 + u_1^2}}, \quad u_1 = \frac{\bar{u}_1}{\sqrt{1 - \bar{u}_1^2}}.
\end{equation}
As $u_1 = (-\infty,0,\infty)$, $\bar{u}_1 = (-1,0,1)$. As it turns out, the compactification of the $q$-direction is trickier because it exists invariant $q$-lines. Then, it has to be compactified by introducing piecewise functions. For completeness, we give the compactification prescription for the entire range of $q$ in detail in appendix \ref{app:compactification}. In this section we only use the compactification of the bottom third region $-\infty<q<\frac{1-\alpha}{\alpha}$. For this region, the $q$-coordinate can be compacted by introducing a new compact variable $\bar{q}$ as
\begin{equation}\label{def_qbar}
\begin{aligned}
& \bar{q} = \frac{1-\alpha}{\alpha} +   \frac{q+1-\frac{1}{\alpha}}{\frac{1}{\alpha}-q}, \; q = \frac{\alpha  (-\alpha
   +\bar{q}+2)-1}{\alpha  (\alpha 
   (\bar{q}+2)-1)}.
\end{aligned}
\end{equation}
As $q = (-\infty,\frac{1-\alpha}{\alpha})$, $\bar{q} = \left(-2+\frac{1}{\alpha },\frac{1-\alpha}{\alpha}\right)$. In terms of the compact dynamical variables $\lbrace\bar{u}_1,\bar{q}\rbrace$, the isotropic vacuum dynamical system \eqref{dynsys_iso_vac} becomes \eqref{B12} and \eqref{B13}. 

The lower bound $\bar{q} = - 2 + \frac{1}{\alpha}$ now represents the locus of all phase space points representing a nonsingular bounce. The physical viability conditions \eqref{phys_viab} and the isotropisation condition \eqref{isotropisation_new} can be written in terms of the compact dynamical variables
\begin{subequations}
\begin{eqnarray}
 && \frac{\bar{u}_1}{j(q(\bar{q})) - q(\bar{q}) - 2} > 0, 
 \label{isotropisation_comp}\\
 && 2 - q(\bar{q}) + \frac{\bar{u}_1}{\sqrt{1 - \bar{u}_1^2}} < 0 
 \label{phys_viab_comp}.
\end{eqnarray}
\end{subequations}
with $j(q)$ and $q(\bar{q})$ coming from Eq. \eqref{ansatz_cosm} and Eq. \eqref{def_qbar} respectively. 

One can note that the dynamical system \eqref{B12} and \eqref{B13} is singular at the lower boundary $\bar{q} = -2 + \frac{1}{\alpha}$, rendering a Jacobian analysis for the equilibrium points on this boundary invalid. Such a problem, however, can be easily regularised using a redefinition of the time variable (see, e.g. Ref. \cite{Bouhmadi-Lopez:2016dzw})
\begin{equation}
    d\tau \rightarrow d\bar{\tau} = \frac{d\tau}{\bar{q} + 2 - \frac{1}{\alpha}} = \frac{\alpha}{\alpha(\bar{q}+2) - 1}d\tau, \label{time-re-definition}
\end{equation}
As we are considering the range $-2+\frac{1}{\alpha} < \bar{q} \leq \frac{1-\alpha}{\alpha}$ here, this time redefinition preserves the arrow of time, and is therefore perfectly viable. In terms of the redefined time variable, the dynamical system becomes
\begin{subequations}\label{dynsys_iso_vac_comp_new}
\begin{align}
 \frac{d\bar{u}_1}{d\bar{\tau}}  = \frac{\sqrt{1-\bar{u}_{1}^2}}{\alpha^2} \Bigg[-2 \left(\alpha ^2 (\bar{q}+1)+\alpha  (\bar{q}+1)-1\right) &
 \nonumber \\
 +\bar{u}_{1}^2 \left(\alpha +\alpha ^2 (3 \bar{q}+4)+2 \alpha  \bar{q}-2\right) & \nonumber \\
 -\sqrt{1-\bar{u}_{1}^2} \bar{u}_{1} \left(3 \alpha ^2+(2 \alpha +1) \alpha  \bar{q}-1\right)\Bigg], &\nonumber\\
\\
 \frac{d\bar{q}}{d\bar{\tau}}  = 2\left(\bar{q}-\frac{1-\alpha}{\alpha}\right)   \left(\bar{q}+2-\frac{1}{\alpha}\right)  \left[\bar{q}\left(\frac{1-\alpha}{\alpha}\right)-1+\frac{3}{\alpha}-\frac{1}{\alpha^2}\right] & . 
\end{align}
\end{subequations}
The equilibrium points of the dynamical system \eqref{dynsys_iso_vac_comp_new} are listed in table \ref{tab:fp_iso_vac}. We complement this information in table \ref{TableXXVIII} in the appendix.

\begin{table}[H]
\resizebox{\textwidth}{!}{
\centering
\begin{tabular}{|c|c|c|c|c|c|}
\hline
Point & Coordinates $(\bar{u}_1,\bar{q})$ & Stability & Cosmology & Physical viability & \multicolumn{1}{|p{4cm}|}{\centering Behaviour of small \\anisotropy} \\
\hline
$\mathcal{K}_{1-}$ & $\left(\frac{\alpha -\sqrt{\alpha  (\alpha +10)+1}+1}{2 \alpha  \sqrt{\frac{\left(\alpha -\sqrt{\alpha  (\alpha
   +10)+1}+1\right)^2}{4 \alpha ^2}+1}}, 
   \frac{1}{\alpha }-1\right)$ & Saddle & $a\sim (-t)^{\alpha}$& $\frac{1}{2}<\alpha<1$ &  \multicolumn{1}{|p{4cm}|}{\centering Decreasing for \\ $0<\alpha<\frac{1}{2}$ \\ Increasing for $\frac{1}{2}<\alpha<1$ \\ Inconclusive for \\ $\alpha= \frac{1}{2}$} \\
\hline
$\mathcal{K}_{1+}$ & $\left(\frac{\alpha +\sqrt{\alpha  (\alpha +10)+1}+1}{2 \alpha  \sqrt{\frac{\left(\alpha +\sqrt{\alpha  (\alpha
   +10)+1}+1\right)^2}{4 \alpha ^2}+1}}, 
   \frac{1}{\alpha }-1\right)$ & Unstable & $a\sim (-t)^{\alpha}$ & $0<\alpha<\frac{1}{2}$ & Increasing for $\alpha>0$ \\
\hline
$\mathcal{C}_{1-}$ & $\left(-1,\frac{1-\alpha}{\alpha} \right)$ & Unstable &  $a\sim (-t)^{\alpha}$& $\frac{1}{2}<\alpha<1$ & Decreasing \\
\hline
$\mathcal{C}_{1+}$ & $\left(1,\frac{1-\alpha}{\alpha} \right)$ & Saddle &  $a\sim (-t)^{\alpha}$& $0<\alpha<\frac{1}{2}$ &  Increasing \\
\hline
$\mathcal{C}_{2-}$ & $\left(-1,-2+\frac{1}{\alpha} \right)$ & Stable & Nonsingular bounce & Viable & Inconclusive \\
\hline
$\mathcal{C}_{2+}$ & $\left(1,-2+\frac{1}{\alpha} \right)$ & Stable & Nonsingular bounce &  Not viable & Increasing \\
\hline
$\mathcal{C}_{3}$ & $\left(-\frac{2}{\sqrt{5}},-2+\frac{1}{\alpha} \right)$ & Saddle & Nonsingular bounce & Viable & Increasing \\
\hline
\end{tabular}}
\caption{Equilibrium points of isotropic vacuum contracting cosmologies in $f(R)$ gravity representing a smooth transition from an ekpyrotic contraction phase to a nonsingular bounce as given by the ansatz \eqref{ansatz}.}
\label{tab:fp_iso_vac}
\end{table}

\begin{figure}
   \centering
    \subfigure[\label{fig:iso_vac_comp_0.2}]{\includegraphics[width=0.45\linewidth]{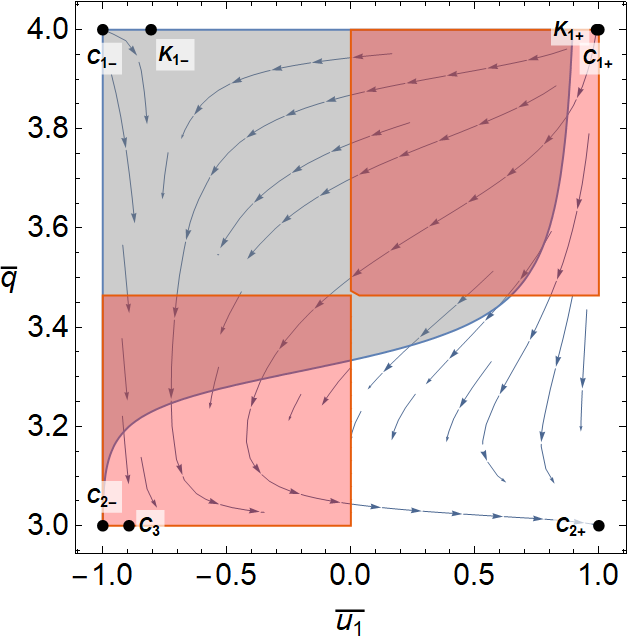}}
    \hspace{0.5cm}
   \subfigure[\label{fig:iso_vac_comp_0.4}]{\includegraphics[width=0.45\linewidth]{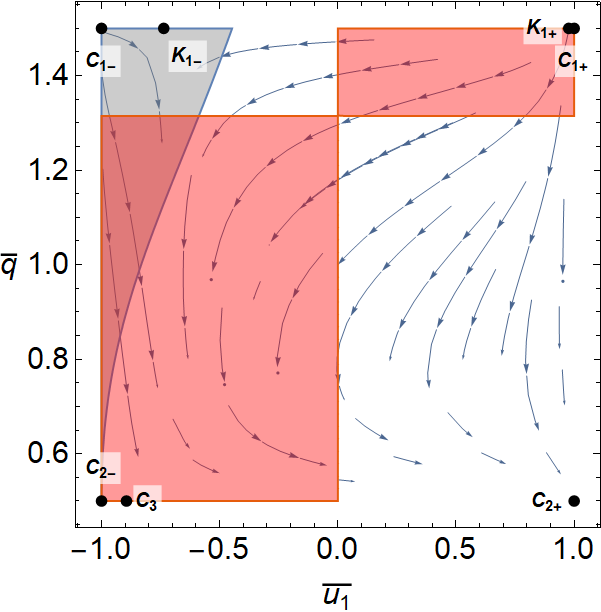}}
    \hspace{0.5cm}
    \subfigure[\label{fig:iso_vac_comp_0.6}]{\includegraphics[width=0.45\linewidth]{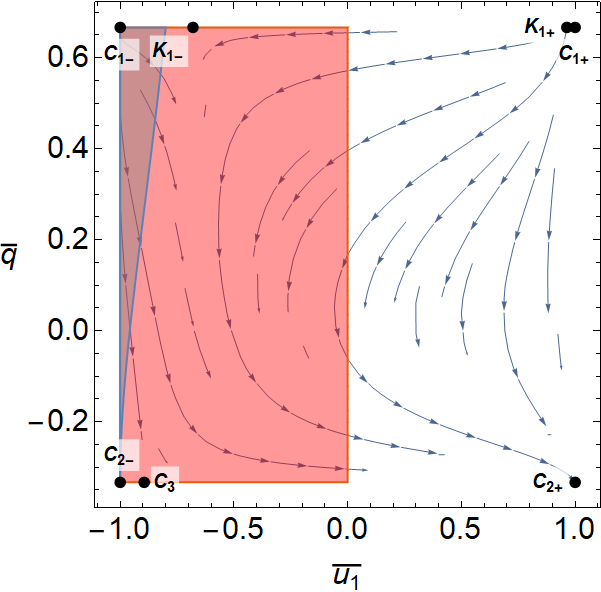}}
    \hspace{0.5cm}
    \subfigure[\label{fig:iso_vac_comp_0.8}]{\includegraphics[width=0.45\linewidth]{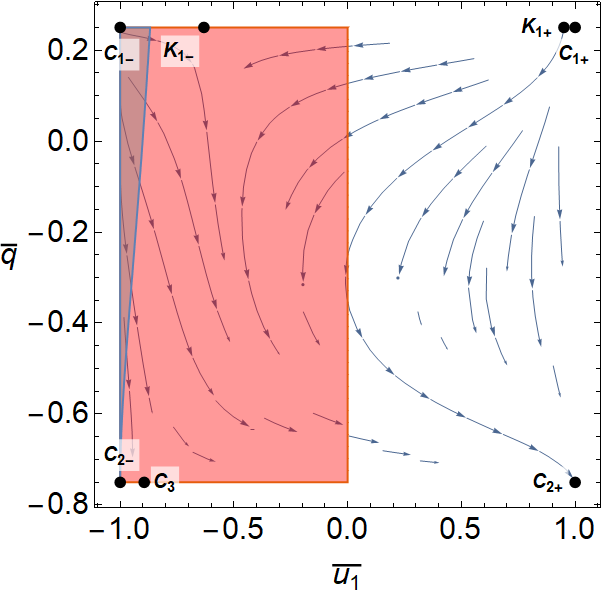}}
    \caption{  \label{fig:iso_vac_comp} The phase portrait on the compactified bottom third region of the isotropic vacuum invariant submanifold as given by the 2-dimensional dynamical system in Eq. \eqref{dynsys_iso_vac_comp_new} for \ref{fig:iso_vac_comp_0.2} $\alpha=0.2$, \ref{fig:iso_vac_comp_0.4} $\alpha=0.4$, \ref{fig:iso_vac_comp_0.6} $\alpha=0.6$, \ref{fig:iso_vac_comp_0.8} $\alpha=0.8$. The pink shaded region corresponds to the region of physical viability as given by the condition \eqref{phys_viab_comp}. The blue shaded region corresponds to the region of isotropisation as given by the condition \eqref{isotropisation_comp}.}
  \end{figure}

We show the 2-dimensional phase portraits for the compactified bottom third region for different values of $\alpha$ in Fig. \ref{fig:iso_vac_comp}.

To translate the equilibrium point $\mathcal{K}_{1-}$ to the origin, and transform the linear part of the dynamical system \eqref{dynsys_iso_vac_comp_new} to its real Jordan form one define the new variables \eqref{translation}, see Appendix \ref{AppC}. 

\begin{figure}
    \centering
    \includegraphics[width=1.0\linewidth]{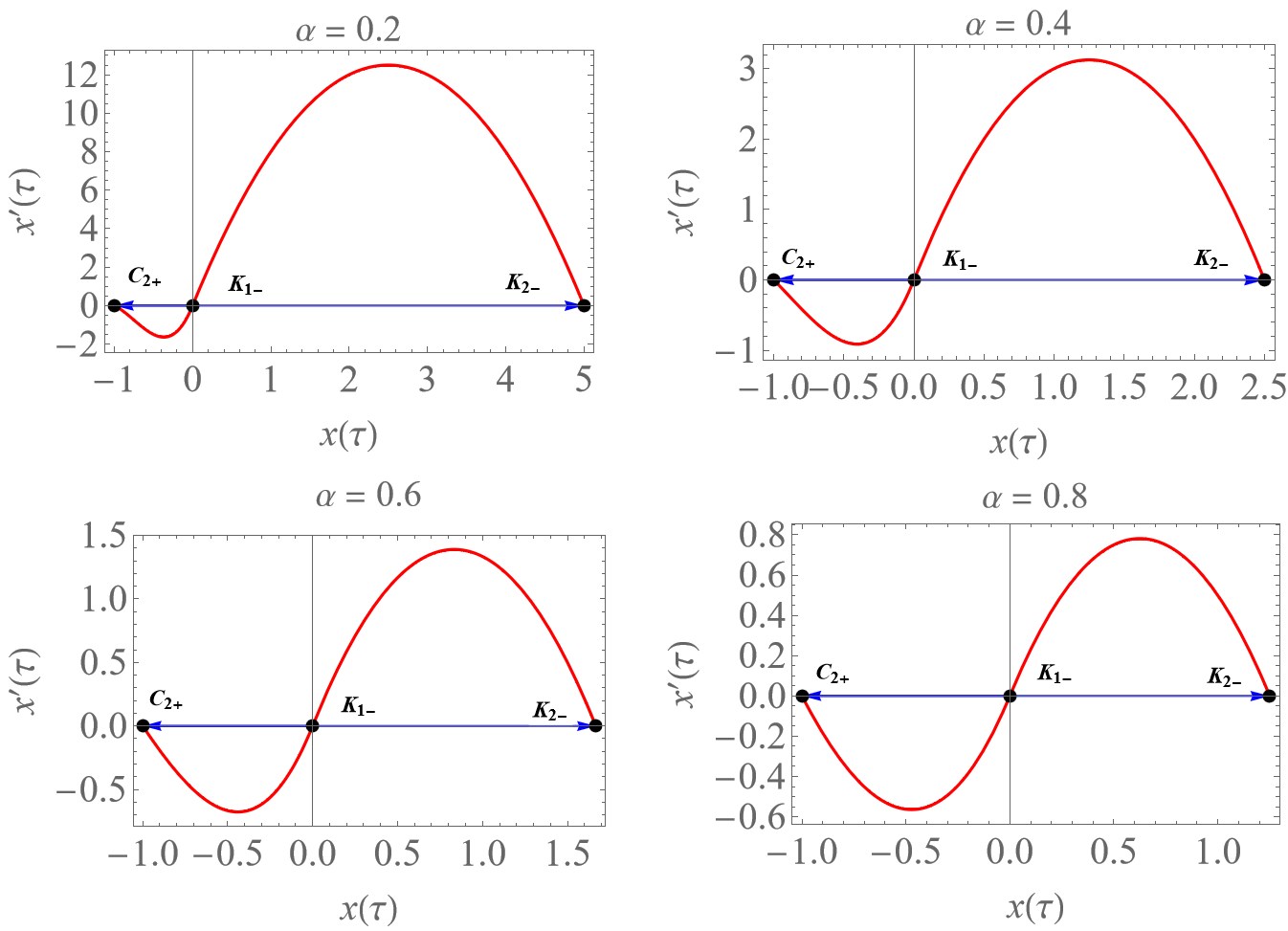}
    \caption{One-dimensional flow of \eqref{syst-1D}. That illustrates that the unstable manifold of  $\mathcal{K}_{1-}$ connects the point $\mathcal{K}_{1-}$ with the late-time attractors $\mathcal{C}_{2 +}$ and $\mathcal{K}_{2 -}$.}
    \label{fig:1Dflow}
\end{figure}

The one-dimensional flow gives the dynamics of the unstable manifold: 
\begin{align}
\label{syst-1D}
      x^{\prime}= \left\{\begin{array}{cc}
        \frac{2 x}{\alpha } + \left(\frac{4}{\alpha }-2\right) x^2 +\left(\frac{2}{\alpha }-2\right) x^3, &  -1 \leq x<0 \\\\
        \frac{2 x}{\alpha }-2 x^2,  & 0<  x \leq \frac{1}{\alpha} 
       \end{array}\right..
\end{align}
The invariant local unstable manifold of  $\mathcal{K}_{1-}$ connects the point $\mathcal{K}_{1-}$ with the late-time attractors $\mathcal{C}_{2 +}$ and $\mathcal{K}_{2 -}$. Then, one can argue for the possible existence of a heteroclinic orbit (sometimes called a heteroclinic connection) as a path in phase space which joins two different equilibrium points (see Fig. \ref{fig:1Dflow}).

\section{Discussion and Conclusion}\label{sec:ending}

In this work, we addressed the important issue of isotropisation during a pre-bounce contracting universe, which is a fundamental issue to address while constructing a nonsingular bouncing paradigm. It is well known that one typically requires a super-stiff fluid \cite{Erickson:2003zm} or a non-ideal fluid \cite{Dunsby:2003sr, Bozza:2009jx} to suppress the anisotropy growth in a contracting universe. The super-stiff fluid is usually modelled via a scalar field fast rolling down a steep negative potential, which gives rise to the ekpyrotic contraction phase \cite{Xue:2011nw, Xue:2013iqy}. We have addressed whether $f(R)$ gravity can isotropise a contracting universe without requiring a super-stiff or a non-ideal fluid. We discover that physically viable isotropic contracting cosmologies with the desired quality of isotropisation (i.e. small perturbative anisotropy dies out) is not possible in $R^n$ ($n>1$) or $R + \alpha R^2$ ($\alpha>0$) gravity, but possible in $\frac{1}{\alpha}e^{\alpha R}$ ($\alpha>0$) gravity, (e.g. the equilibrium points $\mathcal{S}_1,\,\mathcal{S}_4,\,\mathcal{S}_5$). 

One could argue that in these cases, the curvature degree of freedom acts as a super-stiff or non-ideal fluid. Although it is an interesting argument from a physics point of view and is worth investigating, it is not easy to verify this argument as, even if a cosmology (i.e. the form of the scale factor $a(t)$) is provided as in the case of an isotropic equilibrium point, solving for $\sigma(t)$ and $R(t)$ is not at all easy, if not outright impossible. One might need to take recourse to numerical analysis. However, as we saw here, the dynamical system approach is quite helpful for this problem. Even though this approach cannot help us answer the question of whether the curvature d.o.f. behaves as a super-stiff fluid or a non-ideal fluid, it still proves the existence of desired solutions  \footnote{We mention, however, that anisotropy evolution in the presence of an isotropic fluid may be analytically tractable for $R^2$ and $R +\alpha R^2$ gravity \cite{Bhattacharya:2017cbn}. However, the treatment is not as illuminating as the dynamical system approach}.

Even though an isotropising contracting solution is found, there is no guarantee that this will lead to a subsequent bounce. That might not be a big issue as it is common in constructing nonsingular bouncing paradigms to incorporate two scalar d.o.f. \cite{Cai:2013kja}, one responsible for the isotropisation and one responsible for the bounce. Nonetheless, it would be interesting if we could achieve both via one single scalar d.o.f. We have tried the same in Sec. \ref{sec:form-independent}, where we took, as an ansatz, a cosmological evolution representing an ekpyrotic contraction phase smoothly connecting to a nonsingular bounce, namely, Eq. \eqref{ansatz_cosm}. One could, of course, try to reconstruct the $f(R)$ based on such an ansatz \cite{Bamba:2013fha}, but the reconstruction method seldom gives any valuable information about the generic dynamical features of a model. Instead, following the line of Ref. \cite{Chakraborty:2021jku}, we circumvented the reconstruction of $f(R)$ and tried to investigate via the dynamical system approach some generic dynamical features of the model, most notably the absence of ghost and tachyonic instability and the behaviour of small anisotropy. We discover that, unfortunately, at least for the ansatz we have considered, there cannot be a physically viable ekpyrotic contraction phase with isotropisation. Whether or not this is a generic result irrespective of the particular ansatz we had considered remains an important question to be explored.

\section*{Acknowledgement}

 S.A. acknowledges CSIR, Govt. of India, New Delhi, for awarding a Senior Research Fellowship. S.M. acknowledges the Department of Science and Technology (DST), Govt. of India, New Delhi, for awarding the Senior Research Fellowship (File No. DST/INSPIRE Fellowship/2018/IF18D676). S.C. acknowledges the financial assistance provided by the North-West University, South Africa, through the postdoctoral grant NWU PDRF Fund NW.1G01487, as well as the accommodation and financial assistance provided kindly by the Department of Mathematics, BITS-Pilani, Hyderabad Campus.  G. L. was funded by Vicerrectoría de Investigación y Desarrollo Tecnológico (Vridt) at Universidad Católica del Norte through Concurso De Pasantías De Investigación Año 2022, Resolución Vridt N° 040/2022 and through Resolución Vridt N° 054/2022.  P.K.S. acknowledges CSIR, New Delhi, India, for financial support to carry out the Research project [No.03(1454)/19/EMR-II Dt.02/08/2019]. We are very much grateful to the honourable referee and the editor for the illuminating suggestions that have significantly improved our work regarding research quality and presentation.

\begin{appendix}

\section{Jacobian Eigenvalues}\label{app:jacobian_eigenvalues}

In this section we list the Jacobian eigenvalues for all the equilibrium points listed in tables \ref{tab:fp_mon_aniso}, \ref{tab:fp_mon_iso}, \ref{tab:fp_mon_vac}, \ref{tab:fp_quad_aniso}, \ref{tab:fp_quad_iso}, \ref{tab:fp_quad_vac}, \ref{tab:fp_exp_aniso}, \ref{tab:fp_exp_iso}, \ref{tab:fp_exp_vac}, \ref{tab:fp_aniso}, \ref{tab:fp_iso}, \ref{tab:fp_vac}.

\subsection{$f(R)=R^n$ ($n>1$)}

\begin{itemize}
\item Anisotropic fluid: The eigenvalues corresponding to isotropic equilibrium points in the presence of an anisotropic fluid for $R^n$ gravity ($n>1$) are listed in Tab. \ref{tab:XIV}. 
\begin{table}[H]
    \centering
    \begin{tabular}{|c|c|}
   \hline
   Point & Eigenvalues  \\
   \hline
    $\mathcal{P}_1$  & $\lbrace 0, 0, -\frac{4n-5}{n-1}, -2 + 3\omega \rbrace$ \\
    \hline
    $\mathcal{P}_2$ & $\lbrace\frac{4n-5}{n-1}, \frac{4n-5}{n-1},\frac{4n-5}{n-1}, 4- \frac{2}{n-1}+\frac{3}{2n-1}+3 \omega \rbrace $\\
   \hline
   \end{tabular}
    \caption{Jacobian eigenvalues corresponding to isotropic equilibrium points in presence of an anisotropic fluid for $R^n$ gravity ($n>1$) as listed in Tab. \ref{tab:fp_mon_aniso}.}     \label{tab:XIV}
\end{table}
\item Isotropic fluid: the eigenvalues corresponding to isotropic equilibrium points in the presence of an isotropic fluid for $R^n$ gravity ($n>1$) as listed in Tab. \ref{tab:XV}, where 
\begin{equation*}
  \lambda_\pm = \frac{\splitdfrac{-6 n^2 \omega +n (9 \omega +3)-3 \omega  -3 \pm \sqrt{n-1} \Bigg\{4 n^3 (3 \omega +8)^2-4 n^2 \left(54 \omega ^2+165 \omega +152\right)} {+3 n \left(87 \omega ^2+226 \omega +139\right)-81 (\omega +1)^2\Bigg\}^{\frac{1}{2}}}}{4 (n-1) n}.  
\end{equation*}
  \begin{table}[H]
    \centering
    \begin{tabular}{|c|c|}
   \hline
   Point & Eigenvalues  \\
   \hline
    $\mathcal{P}_1$  & $\lbrace 0, -\frac{4n-5}{n-1}, -2 + 3\omega \rbrace$ \\
    \hline
    $\mathcal{P}_2$ & $\lbrace\frac{4 n-5}{n-1}, \frac{4 n-5}{n-1},-\frac{2}{n-1}+\frac{3}{2 n-1}+3 \omega +4 \rbrace $\\
    \hline
    $\mathcal{P}_3$ & $\lbrace-\frac{4 n-3 \omega -3}{n-1}, 2-3 \omega,2-3 \omega \rbrace $\\
   \hline
    $\mathcal{P}_4$ & $\lbrace -\frac{3(-1+\omega(2 n-1))}{2n}, \lambda_{+},\lambda_{-} \rbrace $\\
    \hline
   \end{tabular}
    \caption{Jacobian eigenvalues corresponding to isotropic equilibrium points in presence of an isotropic fluid for $R^n$ gravity ($n>1$) as listed in Tab. \ref{tab:fp_mon_iso}.} 
    \label{tab:XV}
\end{table}
\item Vacuum: the eigenvalues corresponding to isotropic equilibrium points in absence of any fluid for $R^n$ gravity ($n>1$) as listed in Tab. \ref{tab:XVI}. 
\begin{table}[H]
    \centering
    \begin{tabular}{|c|c|}
   \hline
   Point & Eigenvalues  \\
   \hline
    $\mathcal{P}_1$  & $\lbrace 0, -\frac{4n-5}{n-1} \rbrace$ \\
   \hline
    $\mathcal{P}_2$ & $\lbrace\frac{4 n-5}{n-1}, \frac{4 n-5}{n-1}\rbrace $\\
    \hline
   \end{tabular}
    \caption{Jacobian eigenvalues corresponding to isotropic equilibrium points in absence of any fluid for $R^n$ gravity ($n>1$) as listed in Tab. \ref{tab:fp_mon_vac}.} 
    \label{tab:XVI}
\end{table}
\end{itemize}

\subsection{$f(R)= R+\alpha R^{2}$ ($\alpha>0$)}

\begin{itemize}
\item Anisotropic fluid: the eigenvalues of Isotropic equilibrium points for $f(R)= R+\alpha R^{2}$ gravity ($\alpha>0$) in presence of an anisotropic fluid as presented in Tab. \ref{tab:XVII}. 
\begin{table}[H]
    \centering
    \begin{tabular}{|c|c|}
   \hline
   Point & Eigenvalues  \\
   \hline
    $\mathcal{Q}_1$  & $\lbrace 3, 3,3,0, 3(1+\omega) \rbrace$ \\
    \hline
    $\mathcal{Q}_2$ & $\lbrace 5, 5 ,5, -4,  3(1+\omega) \rbrace $\\
   \hline
   \end{tabular}
    \caption{Jacobian eigenvalues corresponding to isotropic equilibrium points in presence of an anisotropic fluid for $R + \alpha R^2$ gravity ($\alpha>0$) as listed in Tab. \ref{tab:fp_quad_aniso}.} 
    \label{tab:XVII}
\end{table}
\item Isotropic fluid: the eigenvalues of Isotropic equilibrium points for $R+\alpha R^{2}$ gravity ($\alpha>0$) in the presence of an isotropic fluid are given in Tab. \ref{tab:XVIII},  where 
 \begin{equation*}
     \beta_\pm = \frac{3}{8}(1-3\omega) \pm \sqrt{6 - \frac{15}{64}(1-3\omega)^2}.
 \end{equation*}
\begin{table}[H]
    \centering
    \begin{tabular}{|c|c|}
   \hline
   Point & Eigenvalues  \\
   \hline
    $\mathcal{Q}_1$  & $\lbrace 3,3,0, 3(1+\omega) \rbrace$ \\
    \hline
    $\mathcal{Q}_2$ & $\lbrace 5, 5, -4,  3(1+\omega) \rbrace $\\
   \hline
    $\mathcal{Q}_3$ & $\lbrace -\frac{3}{2}\left(1+\omega\right), \frac{3}{4}\left(1-3\omega\right),\beta_{+},\beta_{-} \rbrace $\\
   \hline 
\end{tabular}
    \caption{Jacobian eigenvalues corresponding to isotropic equilibrium points in presence of an isotropic fluid for $R + \alpha R^2$ gravity ($\alpha>0$) as listed in Tab. \ref{tab:fp_quad_iso}.} 
    \label{tab:XVIII}
\end{table}
\item Vacuum: the eigenvalues of Isotropic equilibrium points for $R+\alpha R^{2}$ gravity ($\alpha>0$) in a vacuum are presented in  Tab. \ref{tab:XIX}. 
 \begin{table}[H]
    \centering
    \begin{tabular}{|c|c|}
   \hline
   Point & Eigenvalues  \\
   \hline
   $\mathcal{Q}_1$  & $\lbrace 3,3,0 \rbrace$ \\
    \hline
    $\mathcal{Q}_2$ & $\lbrace 5, 5, -4 \rbrace $\\
   \hline
\end{tabular}
    \caption{Jacobian eigenvalues corresponding to isotropic equilibrium points in absence of any fluid for $R + \alpha R^2$ gravity ($\alpha>0$) as listed in Tab. \ref{tab:fp_quad_vac}.} 
    \label{tab:XIX}
\end{table}
\end{itemize}

\subsection{$f(R) = \frac{1}{\alpha} e^{\alpha R}$ ($\alpha>0$)}

\begin{itemize}
 \item Anisotropic fluid:  the eigenvalues of Isotropic equilibrium points for $\frac{1}{\alpha} e^{\alpha R}$ gravity ($\alpha>0$) in presence of an anisotropic fluid are presented in Tab. \ref{tab:XX}. 
 \begin{table}[H]
    \centering
    \begin{tabular}{|c|c|}
   \hline
   Point & Eigenvalues  \\
   \hline
    $\mathcal{S}_1$ & $\lbrace -4, -4,0,0, -2 + 3\omega \rbrace$ \\
    \hline
    $\mathcal{S}_2$ & $\lbrace 4, 4 ,4, 0, 4+3 \omega \rbrace $\\
   \hline
   $\mathcal{S}_3$ & $\lbrace \frac{1}{2} \left(3+\sqrt{17}\right), 3,3, \frac{1}{2} \left(3-\sqrt{17}\right), 3(1+\omega) \rbrace $\\
   \hline
\end{tabular}
    \caption{Jacobian eigenvalues corresponding to isotropic equilibrium points in presence of an anisotropic fluid for $\frac{1}{\alpha}e^{\alpha R}$ gravity ($\alpha>0$) as listed in Tab. \ref{tab:fp_exp_aniso}.} 
    \label{tab:XX}
\end{table}
\item Isotropic fluid: the eigenvalues of Isotropic equilibrium points for $\frac{1}{\alpha} e^{\alpha R}$ gravity ($\alpha>0$) in presence of an isotropic fluid are given in Tab. \ref{tab:XXI}.     
\begin{table}[H]
    \centering
    \begin{tabular}{|c|c|}
   \hline
   Point & Eigenvalues  \\
   \hline
    $\mathcal{S}_1$  & $\lbrace -4, -4, 0, -2 + 3\omega \rbrace$ \\
    \hline
    $\mathcal{S}_2$ & $\lbrace 4, 4, 0, 4+3\omega \rbrace $\\
   \hline
    $\mathcal{S}_3$ & $\lbrace \frac{1}{2} \left(3+\sqrt{17}\right), 3, \frac{1}{2}  \left(3-\sqrt{17}\right), 3(1+\omega) \rbrace $\\
    \hline
     $\mathcal{S}_4$ & $\lbrace -4, -4, 2-3\omega, 2-3\omega \rbrace $\\
     \hline
     $\mathcal{S}_5$ & $\lbrace 4, 0, -4-3\omega, -3\omega \rbrace $\\
     \hline
\end{tabular}
    \caption{Jacobian eigenvalues corresponding to isotropic equilibrium points in presence of an isotropic fluid for $\frac{1}{\alpha}e^{\alpha R}$ gravity ($\alpha>0$) as listed in Tab. \ref{tab:fp_exp_iso}.} 
    \label{tab:XXI}
\end{table}
\item Vacuum: the eigenvalues of Isotropic equilibrium points for $\frac{1}{\alpha} e^{\alpha R}$ gravity ($\alpha>0$) in a vacuum are presented in Tab. \ref{tab:XXII}. 
\begin{table}[H]
    \centering
    \begin{tabular}{|c|c|}
   \hline
   Point & Eigenvalues  \\
   \hline
    $\mathcal{S}_1$  & $\lbrace -4,-4,0 \rbrace$ \\
   \hline
    $\mathcal{S}_2$ & $\lbrace 4, 4, 0 \rbrace $\\
    \hline
      $\mathcal{S}_3$ & $\lbrace \frac{1}{2} \left(3+\sqrt{17}\right), 3, \frac{1}{2}  \left(3-\sqrt{17}\right)\rbrace $\\
    \hline
\end{tabular}
    \caption{Jacobian eigenvalues corresponding to isotropic equilibrium points in absence of any fluid for $\frac{1}{\alpha}e^{\alpha R}$ gravity ($\alpha>0$) as listed in Tab. \ref{tab:fp_exp_vac}.}
    \label{tab:XXII}
\end{table}
\end{itemize}

\subsection{Eigenvalues for form-independent analysis of Sec. \ref{sec:form-independent}}

\begin{itemize}
\item Anisotropic fluid: the eigenvalues of Isotropic equilibrium points in the presence of an anisotropic fluid are presented in Tab. \ref{tab:XXIII}.
\begin{table}[H]
    \centering
    \begin{tabular}{|c|c|}
   \hline
   Point & Eigenvalues  \\
   \hline
    $\mathcal{K}_{1-}$ & $\lbrace \frac{2}{\alpha}, -\frac{\sqrt{\alpha ^2+10 \alpha +1}}{\alpha },\frac{-1+ 7\alpha-\sqrt{\alpha ^2+10 \alpha +1}}{2 \alpha },\frac{-1+ 7\alpha-\sqrt{\alpha ^2+10 \alpha +1}}{2 \alpha },\frac{-3 + 7\alpha-\sqrt{\alpha ^2+10 \alpha +1}+6 \alpha  \omega}{2 \alpha } \rbrace$ \\
    \hline
    $\mathcal{K}_{1+}$ & $\lbrace \frac{2}{\alpha}, \frac{\sqrt{\alpha ^2+10 \alpha +1}}{\alpha },\frac{-1+ 7\alpha+\sqrt{\alpha ^2+10 \alpha +1}}{2 \alpha },\frac{-1+ 7\alpha+\sqrt{\alpha ^2+10 \alpha +1}}{2 \alpha },\frac{-3 + 7\alpha+\sqrt{\alpha ^2+10 \alpha +1}+6 \alpha  \omega}{2 \alpha } \rbrace$\\
   \hline
   $\mathcal{K}_{2-}$ & $\lbrace -\frac{2}{\alpha}, -\frac{\sqrt{\alpha ^2+20 \alpha +4}}{\alpha },\frac{-2+ 7\alpha-\sqrt{\alpha ^2+20 \alpha +4}}{2 \alpha },\frac{-2+ 7\alpha-\sqrt{\alpha ^2+20 \alpha +4}}{2 \alpha },\frac{-6 + 7\alpha-\sqrt{\alpha ^2+20 \alpha +4}+6 \alpha  \omega}{2 \alpha } \rbrace$\\
   \hline
   $\mathcal{K}_{2+}$ & $\lbrace -\frac{2}{\alpha}, \frac{\sqrt{\alpha ^2+20 \alpha +4}}{\alpha },\frac{-2+ 7\alpha+\sqrt{\alpha ^2+20 \alpha +4}}{2 \alpha },\frac{-2+ 7\alpha+\sqrt{\alpha ^2+20 \alpha +4}}{2 \alpha },\frac{-6 + 7\alpha+\sqrt{\alpha ^2+20 \alpha +4}+6 \alpha  \omega}{2 \alpha } \rbrace$\\
   \hline
\end{tabular}
    \caption{Jacobian eigenvalues corresponding to isotropic equilibrium points in the presence of an anisotropic fluid as listed in Tab.\ref{tab:fp_aniso}.} 
    \label{tab:XXIII}
\end{table}
\item Isotropic fluid: the eigenvalues of Isotropic equilibrium points in the presence of an isotropic fluid are presented in Tab. \ref{tab:XXIV}.
  \begin{table}[H]
    \centering
    \begin{tabular}{|c|c|}
   \hline
   Point & Eigenvalues  \\
   \hline
    $\mathcal{K}_{1-}$ & $\lbrace \frac{2}{\alpha}, -\frac{\sqrt{\alpha ^2+10 \alpha +1}}{\alpha },\frac{-1+ 7\alpha-\sqrt{\alpha ^2+10 \alpha +1}}{2 \alpha },\frac{-3 + 7\alpha-\sqrt{\alpha ^2+10 \alpha +1}+6 \alpha  \omega}{2 \alpha } \rbrace$ \\
    \hline
    $\mathcal{K}_{1+}$ & $\lbrace \frac{2}{\alpha}, \frac{\sqrt{\alpha ^2+10 \alpha +1}}{\alpha },\frac{-1+ 7\alpha+\sqrt{\alpha ^2+10 \alpha +1}}{2 \alpha },\frac{-3 + 7\alpha+\sqrt{\alpha ^2+10 \alpha +1}+6 \alpha  \omega}{2 \alpha } \rbrace$\\
   \hline
   $\mathcal{K}_{2-}$ & $\lbrace -\frac{2}{\alpha}, -\frac{\sqrt{\alpha ^2+20 \alpha +4}}{\alpha },\frac{-2+ 7\alpha-\sqrt{\alpha ^2+20 \alpha +4}}{2 \alpha },\frac{-6 + 7\alpha-\sqrt{\alpha ^2+20 \alpha +4}+6 \alpha  \omega}{2 \alpha } \rbrace$\\
   \hline
   $\mathcal{K}_{2+}$ & $\lbrace -\frac{2}{\alpha}, \frac{\sqrt{\alpha ^2+20 \alpha +4}}{\alpha },\frac{-2+ 7\alpha+\sqrt{\alpha ^2+20 \alpha +4}}{2 \alpha },\frac{-6 + 7\alpha+\sqrt{\alpha ^2+20 \alpha +4}+6 \alpha  \omega}{2 \alpha } \rbrace$\\
   \hline
   $\mathcal{K}_{3}$ & $\lbrace \frac{2}{\alpha}, -\frac{3 \alpha  \omega -1}{\alpha },-\frac{\sqrt{\alpha ^2 (\alpha  (\alpha +10)+1) (\omega +1)^2}+\alpha  (\omega +1) (\alpha  (6 \omega +7)-3)}{2 \alpha ^2 (\omega +1)}, \frac{\sqrt{\alpha ^2 (\alpha  (\alpha +10)+1) (\omega +1)^2}-\alpha  (\omega +1) (\alpha  (6 \omega +7)-3)}{2 \alpha ^2 (\omega +1)}  \rbrace$\\
   \hline
     $\mathcal{K}_{4}$ & $\lbrace -\frac{2}{\alpha},-\frac{3 \alpha  \omega -2}{\alpha },\frac{1}{2} \left(\frac{6 \alpha -\frac{\sqrt{\alpha ^2 (\alpha  (\alpha +20)+4) (\omega +1)^2}}{\omega +1}}{\alpha ^2}-6 \omega -7\right), \frac{\sqrt{\alpha ^2 (\alpha  (\alpha +20)+4) (\omega +1)^2}-\alpha  (\omega +1) (\alpha  (6 \omega +7)-6)}{2 \alpha ^2 (\omega +1)}  \rbrace$\\
   \hline
\end{tabular}
    \caption{Jacobian eigenvalues corresponding to isotropic equilibrium points in the presence of an isotropic fluid as listed in Tab.\ref{tab:fp_iso}.} 
    \label{tab:XXIV}
\end{table}   
\item Vacuum: the eigenvalues of Isotropic equilibrium points in a vacuum are presented in Tab. \ref{tab:XXV}.
\begin{table}[H]
    \centering
    \begin{tabular}{|c|c|}
   \hline
   Point & Eigenvalues  \\
   \hline
    $\mathcal{K}_{1-}$ & $\lbrace \frac{2}{\alpha}, -\frac{\sqrt{\alpha ^2+10 \alpha +1}}{\alpha },\frac{-1+ 7\alpha-\sqrt{\alpha ^2+10 \alpha +1}}{2 \alpha } \rbrace$ \\
    \hline
    $\mathcal{K}_{1+}$ & $\lbrace \frac{2}{\alpha}, \frac{\sqrt{\alpha ^2+10 \alpha +1}}{\alpha },\frac{-1+ 7\alpha+\sqrt{\alpha ^2+10 \alpha +1}}{2 \alpha } \rbrace$\\
   \hline
   $\mathcal{K}_{2-}$ & $\lbrace -\frac{2}{\alpha}, -\frac{\sqrt{\alpha ^2+20 \alpha +4}}{\alpha },\frac{-2+ 7\alpha-\sqrt{\alpha ^2+20 \alpha +4}}{2 \alpha }\rbrace$\\
   \hline
   $\mathcal{K}_{2+}$ & $\lbrace -\frac{2}{\alpha}, \frac{\sqrt{\alpha ^2+20 \alpha +4}}{\alpha },\frac{-2+ 7\alpha+\sqrt{\alpha ^2+20 \alpha +4}}{2 \alpha } \rbrace$\\
   \hline
\end{tabular}
    \caption{Jacobian eigenvalues corresponding to isotropic equilibrium points in absence of any fluid as listed in Tab.\ref{tab:fp_vac}.}
    \label{tab:XXV}
\end{table}  
\end{itemize}

\section{Compactifying the isotropic vacuum invariant sub manifold}\label{app:compactification}

This section presents the mathematical procedure to compactify the range of the 2-dimensional phase space given by the dynamical system in Eq. \eqref{dynsys_iso_vac}. This compactification procedure can be used whenever there are more than one invariant submanifolds parallel. The isotropic vacuum invariant submanifold is spanned by $u_1$ and $q$. $u_1$ can be compacted following the usual Poincare compactification
\begin{equation}
\bar{u}_1 = \frac{u_1}{\sqrt{1 + u_1^2}},
\end{equation}
defined in such a way that as $u_1 = (-\infty,0,\infty)$, $\bar{u}_1 = (-1,0,1)$.
 
We want to compactify the $q$-direction in such a way as to preserve the invariant submanifolds to the same values. Let us define
\begin{equation}
   r = q- \frac{1-\alpha}{\alpha}, 
\end{equation}
for $-\infty<q\leq \frac{1-\alpha}{\alpha}$, such that $r\leq0$. Now we use the define the compact variable
\begin{equation}
    R = r/(1-r),
\end{equation}
such that gives $R\in[-1,0]$. The line $q=\frac{1-\alpha}{\alpha}$ corresponds to $r=0$ or $R=0$. Then, we translate the origin to have $\bar{q}=\frac{1-\alpha}{\alpha}$ at $q=\frac{1-\alpha}{\alpha}$. 
Now we define $p=q-  \frac{2-\alpha}{\alpha}$  for $ \frac{2-\alpha}{\alpha}\leq q<\infty$, that gives a positive $p$, then, we use the compactification $p/(1+p)$ that gives a number between $[0,1]$. The  line $q=\frac{2-\alpha}{\alpha}$ corresponds to $p=0$.  Then, we translate the origin to have $\bar{q}=\frac{2-\alpha}{\alpha}$ at $q=\frac{2-\alpha}{\alpha}$.
Finally, we define $\bar{q}$ by $q$ when $\frac{1-\alpha}{\alpha}<q<\frac{2-\alpha}{\alpha}$. In summary, we have
\begin{align}
& \bar{q}(q) = \begin{cases}
             &  \frac{1-\alpha}{\alpha} +   \frac{q+1-\frac{1}{\alpha}}{\frac{1}{\alpha}-q},\,\,\,\text{for $-\infty<q\leq \frac{1-\alpha}{\alpha}$}\\
             &   q,\,\,\,\text{for $\frac{1-\alpha}{\alpha}<q<\frac{2-\alpha}{\alpha}$}\\
             & \frac{2-\alpha}{\alpha} +   \frac{q+1-\frac{2}{\alpha}}{q+2-\frac{2}{\alpha}},\,\,\,\text{for $\frac{2-\alpha}{\alpha}\leq q<+\infty$}
             \end{cases},
\end{align}
 defined in such a way that as  $q=(-\infty,\frac{1-\alpha}{\alpha},\frac{2-\alpha}{\alpha},\infty)$, $\bar{q}=\left(-2+\frac{1}{\alpha },\frac{1-\alpha}{\alpha},\frac{2-\alpha}{\alpha},\frac{2}{\alpha }\right)$.

 We calculate
 \begin{align}
& \bar{q}'(q) = \begin{cases}
             &  \frac{\alpha ^2}{(\alpha  q-1)^2},\,\,\,\text{for $-\infty<q\leq \frac{1-\alpha}{\alpha}$}\\
             &   1,\,\,\,\text{for $\frac{1-\alpha}{\alpha}<q<\frac{2-\alpha}{\alpha}$}\\
             &  \frac{\alpha ^2}{(\alpha  (q+2)-2)^2},\,\,\,\text{for $\frac{2-\alpha}{\alpha}\leq q<+\infty$}
             \end{cases}.
\end{align}
Hence, $\bar{q}$ is $C^1$.

The inverse transformation of $\bar{q}$ is 
\begin{align}
& q(\bar{q}) = \begin{cases}
             & \frac{\alpha  (-\alpha
   +\bar{q}+2)-1}{\alpha  (\alpha 
   (\bar{q}+2)-1)},\,\,\,\text{for $-2 +\frac{1}{\alpha }<\bar{q}\leq \frac{1-\alpha}{\alpha}$}\\
             & \bar{q},\,\,\,\text{for $\frac{1-\alpha}{\alpha}<\bar{q}<\frac{2-\alpha}{\alpha}$}\\
             & -\frac{\alpha  (\alpha +2 (\alpha -1) \bar{q}-4)+4}{\alpha 
   (\alpha  \bar{q}-2)},\,\,\,\text{for $\frac{2-\alpha}{\alpha}\leq \bar{q}<\frac{2}{\alpha}$}
             \end{cases},
\end{align}
with derivative 
\begin{align}
& q'(\bar{q}) = \begin{cases}
             & \frac{\alpha ^2}{(\alpha  (\bar{q}+2)-1)^2},\,\,\,\text{for $-2 +\frac{1}{\alpha }<\bar{q}\leq \frac{1-\alpha}{\alpha}$}\\
             & \bar{q},\,\,\,\text{for $\frac{1-\alpha}{\alpha}<\bar{q}<\frac{2-\alpha}{\alpha}$}\\
             & \frac{\alpha ^2}{(\alpha  \bar{q}-2)^2},\,\,\,\text{for $\frac{2-\alpha}{\alpha}\leq \bar{q}<\frac{2}{\alpha}$}
             \end{cases},
\end{align}
that is also $C^1$. Hence, $q\mapsto \bar{q}$ is a diffeomorphism. \\
The dynamical system in terms of compact variables reduces to the following.\\

\textbf{For $-2+\frac{1}{\alpha} < \bar{q} \leq \frac{1-\alpha}{\alpha}$} (lower strip),
the system is given by 
\begin{align}
\frac{d \bar{u}_1}{d\tau}
= \frac{\sqrt{1-\bar{u}_{1}^2}}{\alpha  (\alpha  (\bar{q}+2)-1)} & \Bigg[-2 \left(\alpha ^2 (\bar{q}+1)+\alpha  (\bar{q}+1)-1\right) 
\nonumber \\
& +\bar{u}_{1}^2 \left(\alpha +\alpha ^2 (3 \bar{q}+4)+2 \alpha  \bar{q}-2\right) \nonumber \\
& -\sqrt{1-\bar{u}_{1}^2} \bar{u}_{1} \left(3 \alpha ^2+(2 \alpha +1) \alpha  \bar{q}-1\right)\Bigg], \label{B12}
\\
\frac{d \bar{q}}{d\tau} 
= 2\left(\bar{q}-\frac{1-\alpha}{\alpha}\right)&\left[\bar{q}\left(\frac{1-\alpha}{\alpha}\right)-1+\frac{3}{\alpha}-\frac{1}{\alpha^2}\right]. \label{B13}
\end{align}
The system  \eqref{B12}, \eqref{B13}   admits the equilibrium points summarised in table \ref{TableXXVIII}. The equilibrium points having $\bar{q}=  \frac{1}{\alpha }+\frac{1}{\alpha -1}-1$ do not satisfy the condition $\bar{q}>-2 +\frac{1}{\alpha}$ for $0<\alpha<1$. Therefore, they are omitted. The equilibrium points having $\bar{q}=-2 +\frac{1}{\alpha}$, are added, and their stability is analysed after implementing the time re-definition \eqref{time-re-definition}. 
\begin{table}[h]
\centering
    \begin{tabular}{|c|c|c|c|}
   \hline
Label & Coordinates $(\bar{u}_1, \bar{q})$ & Eigenvalues &  \\ \hline
$\mathcal{C}_{1-}$& $\left(-1,  \frac{1}{\alpha }-1\right)$ & $\left\{\frac{2}{\alpha }, + \infty\right\}$ & source \\ \hline
$\mathcal{C}_{1+}$ & $\left(1, \frac{1}{\alpha }-1\right)$ & $\left\{\frac{2}{\alpha },  -\infty\right\}$ & saddle \\ \hline
$\mathcal{C}_{2-}$& $\left(-1, \frac{1}{\alpha }-2\right)$ & $\{-2,-2\}$ & sink \\\hline
$\mathcal{C}_{2+}$& $\left(1,  \frac{1}{\alpha }-2\right)$  & $\{-2,-2\}$ & sink \\\hline
$\mathcal{C}_3$ & $\left(-\frac{2}{\sqrt{5}}, \frac{1}{\alpha }-2\right)$ & $\{-2,1\}$ & saddle \\\hline
$\mathcal{K}_{1-}$ & $\left(\frac{\alpha -\sqrt{\alpha  (\alpha +10)+1}+1}{2 \alpha  \sqrt{\frac{\left(\alpha -\sqrt{\alpha  (\alpha
   +10)+1}+1\right)^2}{4 \alpha ^2}+1}}, 
   \frac{1}{\alpha }-1\right)$ &$ \left\{\frac{2}{\alpha }, -\frac{\sqrt{\alpha  (\alpha +10)+1}}{\alpha }\right\}$ & saddle \\ \hline
$\mathcal{K}_{1+}$ & $\left(\frac{\alpha +\sqrt{\alpha  (\alpha +10)+1}+1}{2 \alpha  \sqrt{\frac{\left(\alpha +\sqrt{\alpha  (\alpha
   +10)+1}+1\right)^2}{4 \alpha ^2}+1}}, 
   \frac{1}{\alpha }-1\right)$ & $\left\{\frac{2}{\alpha },\frac{\sqrt{\alpha  (\alpha +10)+1}}{\alpha }\right\}$ & source \\ \hline
\end{tabular}
    \caption{Eigenvalues of the equilibrium points of \eqref{B12}, \eqref{B13}.  For the points with $\bar{u}_1=\pm 1$ and $\bar{q} \neq \frac{1}{\alpha }-2$ the leading order in the second eigenvalue is $-\text{sgn}(\bar{u}_1)/\sqrt{1 - \bar{u}_1^2}$. For the equilibrium points having $\bar{q}=-2 +\frac{1}{\alpha}$, their stability is analysed after implementing the time re-definition \eqref{time-re-definition}. } 
    \label{TableXXVIII}
\end{table} 

\textbf{For $\frac{1-\alpha}{\alpha} < \bar{q} <  \frac{2-\alpha}{\alpha}$} (middle strip), the system is given by 
\begin{eqnarray}
 \frac{d \bar{u}_1}{d\tau}
 &=& -\sqrt{1-\bar{u}_1^2} \left[\bar{q} \left(\bar{u}_1 \left(\sqrt{1-\bar{u}_1^2}-2 \bar{u}_1\right)+2\right)-3 \bar{u}_1^2+2 \sqrt{1-\bar{u}_1^2} \bar{u}_1+2\right] \label{B10}\\
 \frac{d \bar{q}}{d\tau}&=&-2 \left(\bar{q}-\frac{1-\alpha }{\alpha }\right) \left(\bar{q}-\frac{2-\alpha }{\alpha }\right) \label{B11}. 
\end{eqnarray} \\
The system \eqref{B10}, \eqref{B11}   admits the equilibrium points summarised in table \ref{TableXXVII}. 
\begin{table}[h]
\centering
    \begin{tabular}{|c|c|c|c|}
   \hline
Label & Coordinates $(\bar{u}_1, \bar{q})$ & Eigenvalues & Stability \\ \hline
$\mathcal{C}_{1-}$& $ \left( -1,  \frac{1}{\alpha }-1\right)$ & $\left\{\frac{2}{\alpha },   \infty\right\}$ & source \\ \hline
$\mathcal{C}_{1+}$& $ \left( 1,  \frac{1}{\alpha }-1\right)$& $\left\{\frac{2}{\alpha }, -\infty\right\}$ & saddle \\ \hline
$\mathcal{K}_{1-}$ & $\left(\frac{\alpha -\sqrt{\alpha  (\alpha +10)+1}+1}{2 \alpha  \sqrt{\frac{\left(\alpha -\sqrt{\alpha  (\alpha
   +10)+1}+1\right)^2}{4 \alpha ^2}+1}}, 
   \frac{1}{\alpha }-1\right)$ &$ \left\{\frac{2}{\alpha }, -\frac{\sqrt{\alpha  (\alpha +10)+1}}{\alpha }\right\}$ & saddle \\ \hline
$\mathcal{K}_{1+}$ & $\left(\frac{\alpha +\sqrt{\alpha  (\alpha +10)+1}+1}{2 \alpha  \sqrt{\frac{\left(\alpha +\sqrt{\alpha  (\alpha
   +10)+1}+1\right)^2}{4 \alpha ^2}+1}}, 
   \frac{1}{\alpha }-1\right)$ & $\left\{\frac{2}{\alpha },\frac{\sqrt{\alpha  (\alpha +10)+1}}{\alpha }\right\}$ & source \\ \hline
 $\mathcal{K}_{2-}$ & $ \left(\frac{\alpha -\sqrt{\alpha  (\alpha +20)+4}+2}{2 \alpha  \sqrt{\frac{\left(\alpha -\sqrt{\alpha  (\alpha +20)+4}+2\right)^2}{4 \alpha ^2}+1}}, 
   \frac{2}{\alpha }-1\right)$ & $\left\{-\frac{2}{\alpha },-\frac{\sqrt{\alpha  (\alpha +20)+4}}{\alpha }\right\}$ & sink \\\hline
 $\mathcal{K}_{2+}$ & $ \left(\frac{\alpha +\sqrt{\alpha  (\alpha +20)+4}+2}{2 \alpha  \sqrt{\frac{\left(\alpha +\sqrt{\alpha  (\alpha +20)+4}+2\right)^2}{4 \alpha ^2}+1}}, 
   \frac{2}{\alpha }-1\right)$ & $\left\{-\frac{2}{\alpha },\frac{\sqrt{\alpha  (\alpha +20)+4}}{\alpha }\right\}$ & saddle \\\hline
$\mathcal{J}_{-}$ & $ \left( -1,  \frac{2}{\alpha }-1\right)$ & $\left\{-\frac{2}{\alpha }, +\infty\right\}$ & saddle  \\ \hline
$\mathcal{J}_{+}$ & $\left( 1,  \frac{2}{\alpha }-1\right)$ & $\left\{-\frac{2}{\alpha }, -\infty\right\}$ & sink \\ \hline   
\end{tabular}
    \caption{Eigenvalues of the equilibrium points of \eqref{B10}, \eqref{B11}. For the points with $\bar{u}_1=\pm 1$ the leading order in the second eigenvalue is $-\text{sgn}(\bar{u}_1)/\sqrt{1 - \bar{u}_1^2}$.} 
    \label{TableXXVII}
\end{table}

\textbf{For $\frac{2-\alpha}{\alpha} \leq \bar{q} <  \frac{2}{\alpha}$} (upper strip), the system is given by 
\begin{align}
 \frac{d \bar{u}_1}{d\tau} 
= \frac{\sqrt{1-\bar{u}_1^2}}{\alpha  (\alpha  \bar{q}-2)} & \Bigg[2 (\alpha -2) \alpha  (\bar{q}+1)+\sqrt{1-\bar{u}_1^2} \bar{u}_1 \left(\alpha ^2-2 \alpha  \bar{q}+4\right) \nonumber \\
& -\left(\bar{u}_1^2 (2 (\alpha -1) \alpha +(\alpha -4) \alpha  \bar{q}+8)\right)+8\Bigg], \label{B8}\\
 \frac{d \bar{q}}{d\tau}
= 2\left(\bar{q}-\frac{2-\alpha}{\alpha}\right) & \left[\bar{q}\left(\frac{1-\alpha}{\alpha}\right)-1+\frac{2}{\alpha}-\frac{2}{\alpha^2}\right] \label{B9}.
\end{align} 
The system \eqref{B8}, \eqref{B9} admits the equilibrium points summarised in table \ref{TableXXVI}. The equilibrium points having $\bar{q}=   \frac{\alpha }{1-\alpha }+\frac{2}{\alpha }$ do not satisfy the condition $\frac{2-\alpha}{\alpha} \leq \bar{q} <  \frac{2}{\alpha}$ for $0<\alpha<1$. Therefore, they are omitted. The equilibrium points having $\bar{q}= \frac{2}{\alpha}$ are added, and their stability is analysed after implementing the time re-definition  
\begin{equation}
    d\tau \rightarrow d\bar{\tau} =- \frac{\alpha}{\alpha \bar{q}-2}d\tau. \label{time-re-definition-Upper}
\end{equation}
  \begin{table}[H]
\centering  
    \begin{tabular}{|c|c|c|c|}
   \hline
Label & Coordinates $(\bar{u}_1, \bar{q})$ & Eigenvalues & Stability \\ \hline
 $\mathcal{K}_{2-}$ & $ \left(\frac{\alpha -\sqrt{\alpha  (\alpha +20)+4}+2}{2 \alpha  \sqrt{\frac{\left(\alpha -\sqrt{\alpha  (\alpha +20)+4}+2\right)^2}{4 \alpha ^2}+1}}, 
   \frac{2}{\alpha }-1\right)$ & $\left\{-\frac{2}{\alpha },-\frac{\sqrt{\alpha  (\alpha +20)+4}}{\alpha }\right\}$ & sink \\\hline
 $\mathcal{K}_{2+}$& $ \left(\frac{\alpha +\sqrt{\alpha  (\alpha +20)+4}+2}{2 \alpha  \sqrt{\frac{\left(\alpha +\sqrt{\alpha  (\alpha +20)+4}+2\right)^2}{4 \alpha ^2}+1}}, 
   \frac{2}{\alpha }-1\right)$ & $\left\{-\frac{2}{\alpha },\frac{\sqrt{\alpha  (\alpha +20)+4}}{\alpha }\right\}$ & saddle \\\hline
$\mathcal{J}_{-}$ & $\left( -1,  \frac{2}{\alpha }-1\right)$ & $\left\{-\frac{2}{\alpha },  +\infty\right\} $ & saddle\\\hline
$\mathcal{J}_{+}$ & $ \left( 1,  \frac{2}{\alpha }-1\right)$ & $\left\{-\frac{2}{\alpha },  -\infty\right\}$ & sink \\\hline
$\mathcal{I}_{1-}$  & $\left(-1, \frac{2}{\alpha }\right)$ & $\{2,2\}$ & source \\\hline
$\mathcal{I}_{1+}$ & $\left(1, \frac{2}{\alpha }\right)$ & $\{2,2\}$ & source \\\hline
$\mathcal{I}_2$ & $\left(-\frac{2}{\sqrt{5}},  \frac{2}{\alpha }\right)$ & $\{2,-1\}$ & saddle\\\hline
\end{tabular}
    \caption{Eigenvalues of the equilibrium points of \eqref{B8}, \eqref{B9}.  For the points with $\bar{u}_1=\pm 1$ the leading order in the second eigenvalue is $-\text{sgn}(\bar{u}_1)/\sqrt{1 - \bar{u}_1^2}$. For the equilibrium points having $\bar{q}= \frac{2}{\alpha}$, their stability is analysed after implementing the time re-definition  \eqref{time-re-definition-Upper}. } 
        \label{TableXXVI}
\end{table} 

\begin{figure*}
   \centering
    \subfigure[\label{fig:iso_vac_compf_0.2}]{\includegraphics[width=0.45\linewidth]{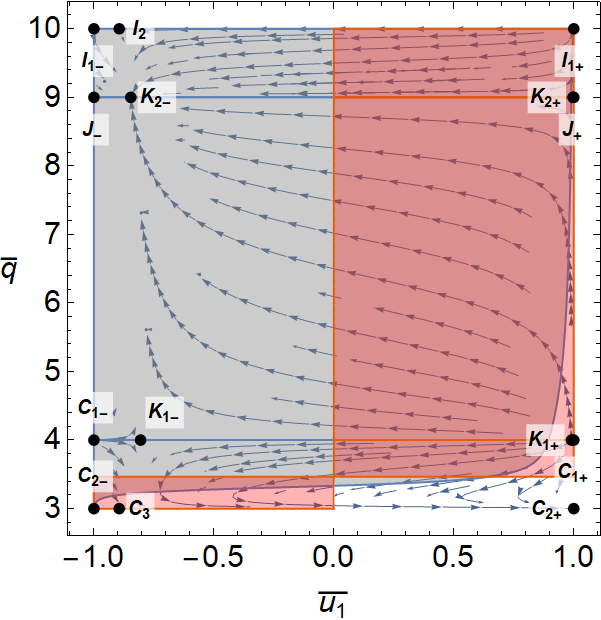}}
    \hspace{0.5cm}
   \subfigure[\label{fig:iso_vac_compf_0.4}]{\includegraphics[width=0.45\linewidth]{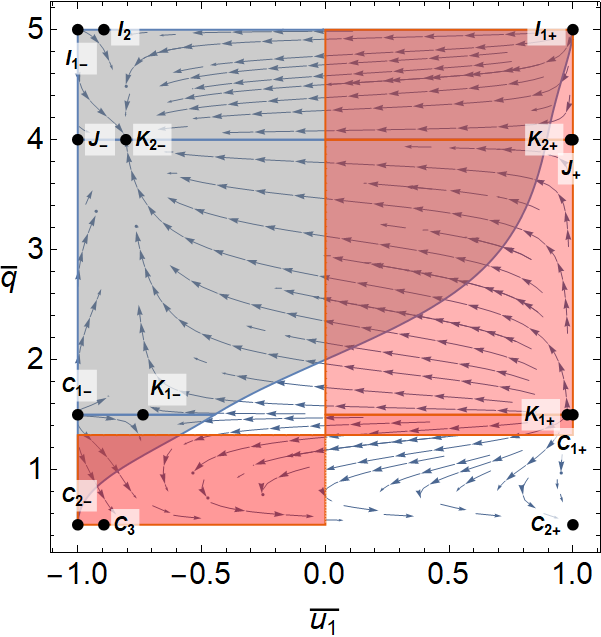}}
    \hspace{0.5cm}
    \subfigure[\label{fig:iso_vac_compf_0.6}]{\includegraphics[width=0.45\linewidth]{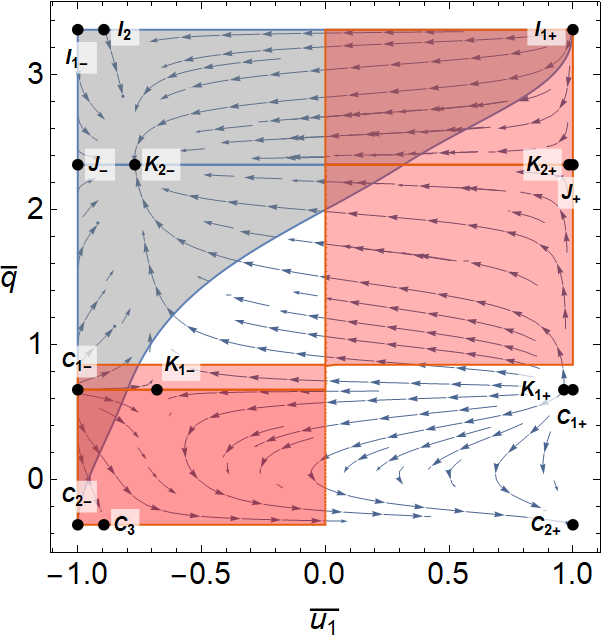}}
    \hspace{0.5cm}
    \subfigure[\label{fig:iso_vac_compf_0.8}]{\includegraphics[width=0.45\linewidth]{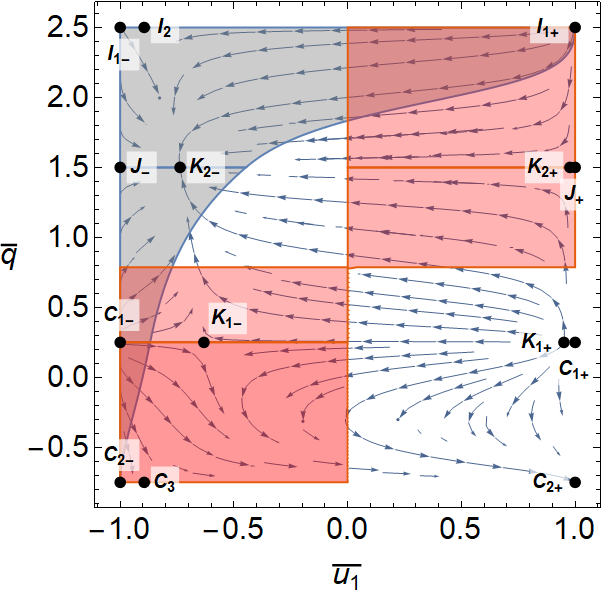}}
    \caption{\label{fig:iso_vac_compf} The phase portrait on the compactified regions of the isotropic vacuum invariant submanifold as given by the 2-dimensional dynamical systems  for \ref{fig:iso_vac_compf_0.2} $\alpha=0.2$, \ref{fig:iso_vac_compf_0.4} $\alpha=0.4$, \ref{fig:iso_vac_compf_0.6} $\alpha=0.6$, \ref{fig:iso_vac_compf_0.8} $\alpha=0.8$.}
    
\end{figure*}

\section{Unstable manifold calculations}
\label{AppC}

Let be the continuous dynamical system described by the ordinary differential equation 
\textbf{\begin{equation}
\mathbf{x}'={\bf X}(\mathbf{x})\label{T1.2}. 
\end{equation}}
Suppose there are equilibria at $\mathbf{x}=\mathbf{x}_0$ and at $\mathbf{x}=\mathbf{x}_1$,  then a solution $\phi(t)$ is a heteroclinic orbit from $\mathbf{x}_0$ and at $\mathbf{x}_1$ if $\phi(t) \rightarrow \mathbf{x}_0$ as $t \rightarrow -\infty$, and  $\phi(t) \rightarrow \mathbf{x}_1$ as $t \rightarrow +\infty$. That implies that the orbit is contained in the stable manifold of $\mathbf{x}_1$  and the unstable manifold of $\mathbf{x}_0$. A heteroclinic cycle is an invariant set in the phase space of a dynamical system. It is a topological circle of equilibrium points and connecting heteroclinic orbits. If a heteroclinic cycle is asymptotically stable, approaching trajectories spend longer periods in a neighbourhood of successive equilibria.

It is well-known that a nonlinear autonomous vector field can be
expressed locally in a neighbourhood of an equilibrium point,
$\bar{\mathbf{x}},$ as

\begin{equation}\label{T1.10}
\mathbf{y}'={\bf A} \mathbf{y}+{\bf R}(\mathbf{y}),\;
\mathbf{y}\in\mathbb{R}^n,
\end{equation}
where ${\bf A}={\bf DX}(\bar{\mathbf{x}}),$ and ${\bf
R}(\mathbf{y})={\cal O}(\|\mathbf{y}\|^2).$

Using elementary algebra \cite{Hirsch} follows that there exists a
lineal transformation, ${\bf T},$ such that the linear part in
\eqref{T1.10}, $\mathbf{y}'={\bf A} \mathbf{y},$ can be expressed
in the real Jordan form
\begin{eqnarray}\label{eq1.10'}
{\bf u}'={\bf A}_s {\bf u}, \quad
{\bf v}'={\bf A}_u {\bf v}, \quad
{\bf w}'={\bf A}_c {\bf w},
\end{eqnarray}
where $${\bf T}^{-1}(\mathbf{y}_1,\mathbf{y}_2,\mathbf{y}_3)\equiv
({\bf u},{\bf v},{\bf w})\in \mathbb{R}^s\times \mathbb{R}^u\times
\mathbb{R}^c,\, s+u+c=n;$$ ${\bf A}_s$ is the $s\times s$ matrix
having eigenvalues with negative real parts; ${\bf A}_u$ is the
$u\times u$ matrix having eigenvalues with positive real parts;
and ${\bf A}_c$ is the $c\times c$ matrix having eigenvalues with
zero real parts. By the change of coordinates induced by ${\bf T}$
the nonlinear vector field \eqref{T1.2} can be expressed as
\begin{eqnarray}\label{eq1.10''}
{\bf u}'={\bf A}_s {\bf u}+{\bf R}_s ({\bf u},{\bf v},{\bf w}), \quad
{\bf v}'={\bf A}_u {\bf v}+{\bf R}_u ({\bf u},{\bf v},{\bf w}), \quad
{\bf w}'={\bf A}_c {\bf w}+{\bf R}_c ({\bf u},{\bf v},{\bf w}),
\end{eqnarray}
where ${\bf R}_s ({\bf u},{\bf v},{\bf w}), {\bf R}_u ({\bf
u},{\bf v},{\bf w}), {\bf R}_c ({\bf u},{\bf v},{\bf w}),$ are,
respectively the first $s, u$ and $c$ components of the vector
field ${\bf T}^{-1}{\bf R}({\bf T y}).$

Let us consider the linear vector field \eqref{eq1.10'}.
Following the previous discussion, the origin of  \eqref{eq1.10'}
have a $s$-dimensional stable invariant manifold; a
$u$-dimensional unstable invariant manifold; and a $c$-dimensional
centre invariant manifold, all of them intersecting the origin.
The following theorem shows how the structure of the invariant
subspaces of the origin change when passing from the study of the
linear system \eqref{eq1.10'} to nonlinear one \eqref{eq1.10''}.

\begin{thm}[Local stable, unstable, and centre manifolds for the origin, theorem 3.2.1 in
\cite{wiggins}]\label{InvMthm}\ If \eqref{eq1.10''}
 is of class $C^r,\; r\geq 2,$ then the equilibrium point $({\bf u},{\bf v},{\bf w})=\mathbf{0}$
 of \eqref{eq1.10''} have a local invariant stable manifold of dimension $s,$
 $W_{\text{loc}}^s(\mathbf{0});$ a local invariant unstable manifold of dimension
 $u,$ $W_{\text{loc}}^u(\mathbf{0});$ and a local invariant centre manifold of dimension
 $c,$ $W_{\text{loc}}^c(\mathbf{0}),$  all of them intersecting at the origin. These manifolds
 are tangent at the origin to the respective invariant subspaces of the linear vector
 field \eqref{eq1.10'}. Then they can be expressed locally as the graphs
\begin{eqnarray}
&& W_{\text{loc}}^s(\mathbf{0})=\left\{({\bf u},{\bf v},{\bf
w})\in \mathbb{R}^s\times \mathbb{R}^u\times \mathbb{R}^c|{\bf
v}={\bf h}_{\bf v}^s({\bf u}), {\bf w}={\bf h}_{\bf w}^s({\bf u}),
\|{\bf u}\|<\delta, \right. \nonumber\\ && \left. {\bf h}_{\bf
v}^s(\mathbf{0})=\mathbf{0},{\bf h}_{\bf
w}^s(\mathbf{0})=\mathbf{0},{\bf Dh}_{\bf
v}^s(\mathbf{0})=\mathbf{0},{\bf Dh}_{\bf
w}^s(\mathbf{0})=\mathbf{0}
\right\};\nonumber\\
&& W_{\text{loc}}^u(\mathbf{0})=\left\{({\bf u},{\bf v},{\bf
w})\in \mathbb{R}^s\times \mathbb{R}^u\times \mathbb{R}^c|{\bf
u}={\bf h}_{\bf u}^u({\bf v}), {\bf w}={\bf h}_{\bf w}^u({\bf v}),
\|{\bf v}\|<\delta, \right. \nonumber\\ && \left. {\bf h}_{\bf
u}^u(\mathbf{0})=\mathbf{0},{\bf h}_{\bf
w}^u(\mathbf{0})=\mathbf{0},{\bf Dh}_{\bf
u}^u(\mathbf{0})=\mathbf{0},{\bf Dh}_{\bf
w}^u(\mathbf{0})=\mathbf{0}
\right\};\nonumber\\
&& W_{\text{loc}}^c(\mathbf{0})=\left\{({\bf u},{\bf v},{\bf
w})\in \mathbb{R}^s\times \mathbb{R}^u\times \mathbb{R}^c|{\bf
u}={\bf h}_{\bf u}^c({\bf w}), {\bf v}={\bf h}_{\bf v}^c({\bf w}),
\|{\bf w}\|<\delta, \right. \nonumber\\ && \left. {\bf h}_{\bf
u}^c(\mathbf{0})=\mathbf{0},{\bf h}_{\bf
v}^c(\mathbf{0})=\mathbf{0},{\bf Dh}_{\bf
u}^c(\mathbf{0})=\mathbf{0},{\bf Dh}_{\bf
v}^c(\mathbf{0})=\mathbf{0} \right\},
\end{eqnarray}
where the functions ${\bf h}_{\bf v}^s,{\bf h}_{\bf w}^s,{\bf
h}_{\bf u}^u,{\bf h}_{\bf w}^u,{\bf h}_{\bf v}^c,$ and ${\bf
h}_{\bf v}^c$ are $C^r$-functions and $\delta$ a positive small
enough number. The orbits at $W_{\text{loc}}^s(\mathbf{0})$ and at
$W_{\text{loc}}^s(\mathbf{0})$ have the same asymptotic properties
as the orbits in the invariant subsets $E^s$ and $E^u$
respectively. That is, the orbits of \eqref{eq1.10''} with initial
conditions at $W_{\text{loc}}^s(\mathbf{0})$ (resp.,
$W_{\text{loc}}^u(\mathbf{0})$) tends asymptotically to the origin
at an exponential rate as $\tau\rightarrow +\infty$ (resp.,
$\tau\rightarrow -\infty$).
 \end{thm}
The conditions ${\bf Dh}_{\bf v}^s(\mathbf{0})=\mathbf{0},{\bf
Dh}_{\bf w}^s(\mathbf{0})=\mathbf{0}, \ldots$ reflect the fact
that the nonlinear manifolds are tangent to the associated
invariant linear subspaces at the origin. In the formulation of theorem \ref{InvMthm}, in expressions like
``local invariant stable manifold $\ldots$'', the term ``local''
is referred to the fact that the manifolds are defined as a graph
only in a small neighbourhood of the equilibrium point. Consequently,
all these invariant manifolds have a boundary. Hence, they are
only locally invariant in that the orbits initially on
them can abandon the local manifold but only cross the
boundary. The invariance maintains because the vector field is
tangent to the manifolds. In case the equilibrium point is hyperbolic (i.e.,
$E^c=\emptyset$), the interpretation of theorem \ref{InvMthm} is
that the trajectories of the nonlinear vector field have
qualitatively the same behaviour as the orbits of the associated linear problem in a neighbourhood of the equilibrium point. 
The stable and unstable manifolds are unique.   Due
to the non-hyperbolicity, the analysis is more difficult for the centre manifold, and in
general, the centre manifold is not unique. However, the centre
manifold is unique in all the orders in its Taylor expansion. All possible invariant manifolds differ only on small
exponential perturbations depending on the distance from the
origin to the equilibrium point (see \cite{wiggins}). It is important to note, however, that unlike the case of a linear
system, the centre manifold, $W_{\text{loc}}^c(\mathbf{0})$ will
contain all those dynamics not classified by linearisation (i.e.,
the non-hyperbolic directions). In particular, this manifold may
contain stable, unstable or neutral regions. The
classification of the dynamics in this manifold can only be
determined by utilising more sophisticated methods, such as centre
manifold theorems or the theory of normal forms (see
\cite{wiggins}).

This section investigates the unstable manifold of  $\mathcal{K}_{1-}$. To translate the equilibrium point $\mathcal{K}_{1-}$ to the origin and transform the linear part of the dynamical system \eqref{dynsys_iso_vac_comp_new} to its real Jordan form, one defines the new variables
\begin{align}\label{translation}
    x & =\bar{q}-\frac{1}{\alpha }+ 1,\quad 
    y   =\bar{u}_1-\frac{\alpha -\sqrt{\alpha  (\alpha +10)+1}+1}{2\alpha \sqrt{\frac{\left(\alpha
   -\sqrt{\alpha  (\alpha +10)+1}+1\right)^2}{4 \alpha ^2}+1}  } + A \left(\bar{q}-\frac{1}{\alpha }+ 1\right),
\end{align}
where the constant $A$ is conveniently chosen. The new system for $-1\leq x<0$ can be symbolically written as 
\begin{align}
    \frac{d x}{d\bar{\tau}} & =\frac{2 x}{\alpha } + \left(\frac{4}{\alpha }-2\right) x^2 +\left(\frac{2}{\alpha }-2\right) x^3, \\
    \frac{d y}{d\bar{\tau}} & =  -\frac{\sqrt{\alpha  (\alpha +10)+1}}{\alpha } y+  g_{-}(x, y), \quad -1\leq  x<0,
\end{align}
where $g_{-}(x, y)$ denotes higher order terms in $x$ and $y$. By continuity of the flow, the dynamics for $x>0$ is obtained after the transformation of system  \eqref{B10}, \eqref{B11} under the change of variables \eqref{translation}; symbolically written as 
\begin{equation}
    \frac{d x}{d \tau}= \frac{2 x}{\alpha }-2 x^2, \quad     \frac{d y}{d\tau}= -\frac{\sqrt{\alpha  (\alpha +10)+1}}{\alpha } y + g_{+}(x, y), \quad 0<x\leq \frac{1}{\alpha}.
\end{equation}
where $g_{+}(x, y)$ denotes higher order terms in $x$ and $y$. 
Observe that the time change $\bar{\tau} \mapsto \tau$ does not affect the orbits. 
Since
$\frac{d x}{d \bar{\tau}}= \frac{2}{\alpha} x + \text{higher order terms}$, it follows that the local unstable manifold of the origin $(x,y)=(0,0)$ is given by the graph 
\begin{equation}
W_{\text{loc}}^u(\mathbf{0}):=    \left\{(x,y) \in \mathbb{R}^2: y=h(x), h(0)=0, h'(0)=0, |x|<\delta\right\}, 
\end{equation}
for an small enough $\delta>0$. From the invariance, and tangentially conditions of $W_{\text{loc}}^u(\mathbf{0})$, the function $h(x)$ satisfies
\begin{align}
&  -\frac{\sqrt{\alpha  (\alpha +10)+1}}{\alpha } h(x) + g_{-}(x,h(x))\nonumber \\
&  -h'(x) \left[\frac{2 x}{\alpha }  + \left(\frac{4}{\alpha }-2\right) x^2 +\left(\frac{2}{\alpha }-2\right) x^3\right]=0, \quad -1\leq  x<0, \label{negative-x}\\
&   -\frac{\sqrt{\alpha  (\alpha +10)+1}}{\alpha } h(x) + g_{+}(x,h(x))-h'(x) \left[ \frac{2 x}{\alpha }-2 x^2\right]=0, \quad 0< x\leq\frac{1}{\alpha}. \label{positive-x}
\end{align}
Taking Taylor expansions, we define 
\begin{equation}
    h(x)= \left\{ \begin{array}{cc}
       a_2 x^2 + a_3 x^3 + a_4 x^4 + \ldots a_{n} x^n + \ldots ,  & -1\leq  x<0 \\
       b_2 x^2 + b_3 x^3 + b_4 x^4 + \ldots b_{n} x^n + \ldots ,  & 0<x\leq \frac{1}{\alpha} \\
    \end{array}\right.,
\end{equation}
where the coefficients $A$,  $a_i, b_i$ are undetermined constants. 
The value of $A$ is chosen such that the linear term in $x$ be zero for all $x$, say,
\begin{align}
    A & = -\frac{12 \sqrt{2}}{\left(\sqrt{\alpha  (\alpha +10)+1}+2\right) \left(-7 \alpha +\sqrt{\alpha 
   (\alpha +10)+1}-11\right)} \nonumber \\
   & \times\frac{ \alpha ^3}{ \sqrt{\alpha  \left(3 \alpha -\sqrt{\alpha  (\alpha +10)+1}+6\right)-\sqrt{\alpha 
   (\alpha +10)+1}+1}}.
\end{align}

Then,  $a_i, b_i$ are calculated order by order, but equating to zero the coefficients of the same power in \eqref{negative-x} for negative $x$ and in \eqref{positive-x} for positive $x$. Combining altogether, in Fig. \ref{fig:iso_vac_compf} is presented the phase portrait on the compactified regions of the isotropic vacuum invariant submanifold as given by the 2-dimensional dynamical systems  for \ref{fig:iso_vac_compf_0.2} $\alpha=0.2$, \ref{fig:iso_vac_compf_0.4} $\alpha=0.4$, \ref{fig:iso_vac_compf_0.6} $\alpha=0.6$, \ref{fig:iso_vac_compf_0.8} $\alpha=0.8$. Hence, one can find a global representation of the phase portrait in compact variables. This formulation enabled us to understand the global structure of the reduced phase space corresponding to the isotropic vacuum invariant sub-manifold. 

\end{appendix}

\bibliographystyle{unsrtnat}
\bibliography{refs.bib}
\end{document}